# Numerical Modeling of a Metallic Alloy Droplet Impact, Solidification, and Macrosegregation

Ehsan Amani* and Ali Moeinirad

*Department of Mechanical Engineering, Amirkabir University of Technology (Tehran Polytechnic), Iran*

**Abstract**

For the first time, we study the effects of mass transfer on alloy droplet impact, solidification, and solute non-homogeneity or MacroSegregation (MS). For this purpose, a multi-physics model, including fully-coupled hydrodynamics of multiphase flow, wetting, conjugate heat transfer, phase change, and mass transfer, is established and validated carefully across several benchmarks. In addition, a concentration-flux-correction algorithm is proposed to conserve the solute mass within the droplet at the numerical discretization level. The validated model is subsequently used to explore the impact and solidification of an alloy droplet. The results show that accounting for mass transfer and solute transport improves the prediction of the dynamics of a solidifying splat, in a consistent manner with experimental observations, while this effect has been neglected in all previous studies. The analysis demonstrates that this improvement results from the flow sensitivity to the mushy region thickness, which is regulated by local solute concentration through the solidus and liquidus temperatures. Furthermore, analyses of the MS map identify several critical segregation regions most susceptible to defect initiation, such as cracking, including a long negatively segregated vein within the lamella, formed by the severe advective flow through the mush in the middle stages of the impact, a positively segregated region near the rim of the splat, appearing as a result of hook formation, and a major positively segregated zone at the top of the rim, which develops following the accumulation of the solute-rich melt sustained by washout from the lamella.

* Corresponding author. Address: Mechanical Engineering Dept., Amirkabir University of Technology (Tehran Polytechnic), 424 Hafez Avenue, Tehran, P.O.Box: 15875-4413, Iran. Tel: +98 21 64543404. Emails: eamani@aut.ac.ir (E. Amani)

## 1. Introduction

The impact of a liquid droplet onto a solid surface is a multi-physics problem of significant scientific and industrial importance. The quality, resolution, and integrity of products in applications such as ink-jet printing, thermal spray coating, spray-cooling, and laser metal deposition, are directly dictated by the behavior of individual droplets upon striking a substrate. In the automotive sector, understanding spray-wall impingement in port-fuel internal combustion engines is critical for efficiency and emissions control. When a droplet is composed of a molten metallic alloy, the process becomes substantially more complex, involving the coupled interplay of fluid hydrodynamics, surface wetting, heat transfer, solidification phase change, and mass transfer. A particularly relevant application is additive manufacturing, especially droplet-based metal 3D printing processes.

The hydrodynamics of droplet-wall impact is primarily governed by the interplay of inertial, viscous, and capillary forces. This is typically characterized by dimensionless parameters, most notably the Weber number (We), representing the ratio of inertia to capillary force, and the Reynolds number (Re), which represents the ratio of inertia to viscous force (Josserand & Thoroddsen 2016). Furthermore, the wettability of a droplet on the solid surface, often quantified by the macroscopic Contact Angle (CA) and Contact Angle Hysteresis (CAH), plays a crucial role in the spreading and retraction dynamics (Riboux & Gordillo 2014, Josserand & Thoroddsen 2016). Rioboo, *et al.* (2001, 2002) have shown that the impact of a liquid droplet on a dry wall can result in at least six distinct regimes, including deposition, prompt splash, corona splash, receding breakup, partial rebound, and complete rebound, with a specific outcome being highly dependent on We and surface wetting. Other complex regimes, such as lifted thin film (Palacios *et al.* 2013), deposition with central thin jet eruption (Siddique *et al.* 2020), and fingering-rebound (Rioboo *et al.* 2002), have also been documented in the literature.

Modeling the droplet-wall hydrodynamics usually requires the use of a Fully-Resolved Simulation (FRS) multiphase flow model with an appropriate wetting boundary condition at the solid surface, accounting for Contact-Line (CL) dynamics (Sui *et al.* 2014). To capture interface evolution, the Volume-Of-Fluid (VOF) method with a Piecewise Linear Interface Calculation (PLIC) interface reconstruction (Tryggvason *et al.* 2011) has been used in the majority of the studies on droplet-wall impact, see e.g., (Bussmann *et al.* 2000, Fujimoto *et al.* 2007, Lunkad *et al.* 2007, Zhang *et al.* 2016a, Yang *et al.* 2018), while other FRS techniques have also been successfully implemented for the impact problem, like Level-Set (LS) (Ding & Spelt 2007, Caviezel *et al.* 2008, Pan *et al.* 2022), Front-Tracking Method (FTM) (Muradoglu & Tasoglu 2010), and Phase-Field (PF) (Zhang *et al.* 2016b). Amani and Abdi-Sij (2024) conducted a comparative study on VOF-based models, using different algebraic and geometric interface reconstructions, over a wide range of impact We and surface conditions. They pointed out the higher accuracy of the hybrid Combined LS-VOF (CLSVOF) compared to the others. In terms of CL-dynamics, several comparative studies investigated the performance of different CA closures (Malgarinos *et al.* 2014, Amani & Abdi-Sij 2024). Recently, to cover a wider range of conditions, Kawakami *et al.* (2023) proposed a hybrid low-speed-slipping/high-speed-rolling model, using a modified variant of the relaxed impermeability condition.

For a molten droplet impact and solidification, the process is further complicated by heat transfer and phase change. The coupled interaction between the impact dynamics and the advancing solidification front is a key feature of this process (Attinger *et al.* 2000). In terms of modeling, the problem is a three-phase flow with three pairs of interfaces between the molten droplet, solidified droplet, and surrounding gas. This group of problems can be divided into two categories: The pure-material solidification, where all these interfaces are physically sharp, and the alloy solidification, where the region between the pure melt and completely solidified droplet can macroscopically be considered as a thick porous layer, called "mushy" region. For the pure-material droplet impact and solidification, the liquid-solid interface can also be modeled using a sharp interface approach. In a pioneering work, for a tin droplet impinging upon a steel plate, Pasandideh-Fard *et al.* (2002) combined the VOF-PLIC method for droplet-gas interface with the fixed velocity method for the solidification front. Alternatively, many researchers adopted the well-known Enthalpy-Porosity (EP) model for the solidification front. This method was originally proposed for alloy solidification, involving a thick mushy region (Voller & Prakash 1987). However, for a pure material with a sharp solidification front, a very small phase-change (fictitious) temperature difference can be chosen to approximate a constant-temperature solidification. Through a sensitivity analysis, Zhang *et al.* (2018) showed that, for water droplet icing, the results show a negligible dependence on this assumption when the fictious temperature difference is below 1℃. In a subsequent study (Zhang *et al.* 2020b), they calibrated the mushy region constant, governing the velocity damping. For the same application, Shinan *et al.* (2019) used CLSVOF for the droplet-gas interfaces and EP for the solidification front. On the other hand, Shen *et al.* (2019) combined the PF with EP method for these interfaces, respectively. Some other studies modeled all three phases of the problem by a unified approach, such as FTM (Vu *et al.* 2015), Lagrangian sub-mesh VOF (Vincent *et al.* 2015), Lattice-Boltzmann (Zhang *et al.* 2020a), and meshless explicit moving particle simulation (Fukudome *et al.* 2021).

In case of alloy-droplet solidification, EP is the most straightforward choice to treat the thick mushy layer, and has been used by Amani and Sharifi (2024, 2025) along with a CLSVOF method for (molten and solidified) droplet-gas interfaces in 7075 aluminum alloy 3D printing. However, they did not account for the effect of alloy component concentration transport and solute redistribution within the droplet during solidification, which introduces compositional inhomogeneity. This inhomogeneity of alloy components after complete solidification, called "MacroSegregation" (MS) (Stefanescu 2015), has a negative impact on the quality of the solidified object. Several factors can affect MS in alloy solidification (Pickering 2013). The physical mechanisms driving MS are diverse and include the relative motion between the liquid and solid phases, fluid flow within the semi-solid mushy zone arising from natural convection, and liquid-phase shrinkage (Beckermann 2002). Here a brief review of macro-scale and continuum mechanics modeling of MS, relevant to this work, is provided. For more detailed surveys, including meso-scale and micro-scale models, readers can consult references (Pickering 2013, Stefanescu 2015, Dong *et al.* 2017).

The modeling of MS began with the pioneering work of Flemings and Nereo (1967), who developed an analytical model for inter-dendritic flow. Subsequent models were based on either one-fluid Enthalpy-Porosity (EP) approaches, such as the continuum model developed by Bennon and Incropera (1987a), or volume-averaged two-phase models (Ni & Beckermann 1991). More advanced frameworks have been developed to account for phenomena like the Columnar-to-Equiaxed Transition (CET) by extending the models to three or even five phases (Wu & Ludwig 2006, Wu *et al.* 2010, Zheng *et al.* 2016, Wang *et al.* 2019). However, a key challenge in these models is accurately representing the fluid dynamics within the mushy zone, which can behave as a slurry of freely moving equiaxed crystals or as a porous medium of rigid columnar dendrites. To address this, hybrid permeability models were introduced, which switch between different physical laws. For example, Chang and Stefanescu (1996) developed an EP-based hybrid model that accounts for transition from a pseudo-viscosity in the slurry region to Darcy's law in the porous region, at a critical solid fraction. Recently, Moeinirad and Amani (2025) proposed a modification to such a hybrid model to ensure a continuous and more physically consistent transition of characteristics between the slurry and porous regions, which was shown to improve the prediction of key features in the final MS map.

Although a well-established topic in casting applications, the species transport effects on the alloy-droplet impact and solidification have not been addressed in the literature. This effect is also crucial to predict MS in alloy 3D printing applications. The primary objective of the present research is to fill this gap by implementing a detailed numerical framework within the open-source software OpenFOAM (www.openfoam.org). This requires the integration of species transport and microsegregation models into a multiphase flow model, with a technique to prevent numerical species diffusion and leakage at the gas-droplet interface. This is a challenging task which is dealt with carefully in the present study. Owing to the novelty of the full hydrodynamics, wetting, heat and mass transfer, and phase-change modeling of an alloy-droplet impact and solidification, all governing dimensionless parameters are introduced, including several new ones presented in this study for the first time. All these matters are addressed in section 2 after a problem statement. In section 3, the details of the numerical solution procedure are reported. In section 4, the assessment of each sub-model of the proposed modeling framework is systematically performed against a series of benchmarks of increasing complexity. Then, the model is used to investigate the impact, solidification, and MS map features of an alloy droplet. Finally, the major conclusions of the study are summarized in section 5.

## 2. Physical and mathematical modelling

### *2.1. Problem statement*

Here, the impact, solidification, and MS of a single droplet of A380 aluminum alloy (AlA380) is of main interest. The specific values of parameters are taken from the experiment by Xue *et al.* (2007). The problem involves a spherical molten AlA380 droplet of diameter $D_0$ $(= 3.92\ mm)$ impacting upon a steel solid wall (substrate) in an air environment. The initial temperatures of the droplet, wall, and air are $T_0$ $(= 903\ K)$, $T_{w,0}$ $(= 473\ K)$, and $T_{g,0}$ $(= 298\ K)$, respectively. The impact velocity is $V_0$ $(= 3\ ms^{-1})$. The substrate surface is characterized by a CA of $\theta_c$ $(= 90°)$ and thermal contact resistance of $R_c'' = 1.64 \times 10^{-6}\ m^2KW^{-1}$. The nominal composition of AlA380 is 85.5% Al, 8.5% Si, 4% Cu, and 2% Fe. However, we use a binary Al–Si alloy approximation, including the dominant components, with a silicon concentration of 8.5wt%. The thermophysical properties and phase diagram data for the AlA380 alloy, air, and steel substrate are given in Table 4 and Table 5 of Appendix A. The binary alloy assumption remains widely used in MS simulations because it provides a computationally efficient and physically transparent approach for capturing the dominant features of solidification. The binary models can still reproduce the primary mechanisms governing MS, including solute rejection, thermosolutal convection, and mushy-zone transport. Several studies showed that, in cases where one alloy component dominates buoyancy and partitioning behavior, binary approximations can successfully predict overall segregation trends and even yield results close to more complex multicomponent models, especially for global flow and macro-scale solute redistribution (Schneider & Beckermann 1995, Mo & Thevik 1998). However, it should be noted that while binary models often capture the general structure of MS, they can miss quantitative details such as the influence of minor alloy components on density-driven flow and phase stability.

Assuming an axisymmetric problem, the computational domain is illustrated in Figure 1a which is composed of two rectangular regions (a multi-region, multiphase flow problem). The upper region is the multiphase flow region and resolves interactions among the molten droplet, the solidified layer, and the surrounding gas, while the lower region is dedicated to heat conduction within the solid substrate. The dimensions of the multiphase flow domain are selected through a trial-and-error process to eliminate domain-size effects, resulting in a radius of $l$ ($= 7.5\ mm$), a height of $H_a$ ($= 12\ mm$), and substrate thickness of $H_s$ ($= 2\ mm$).

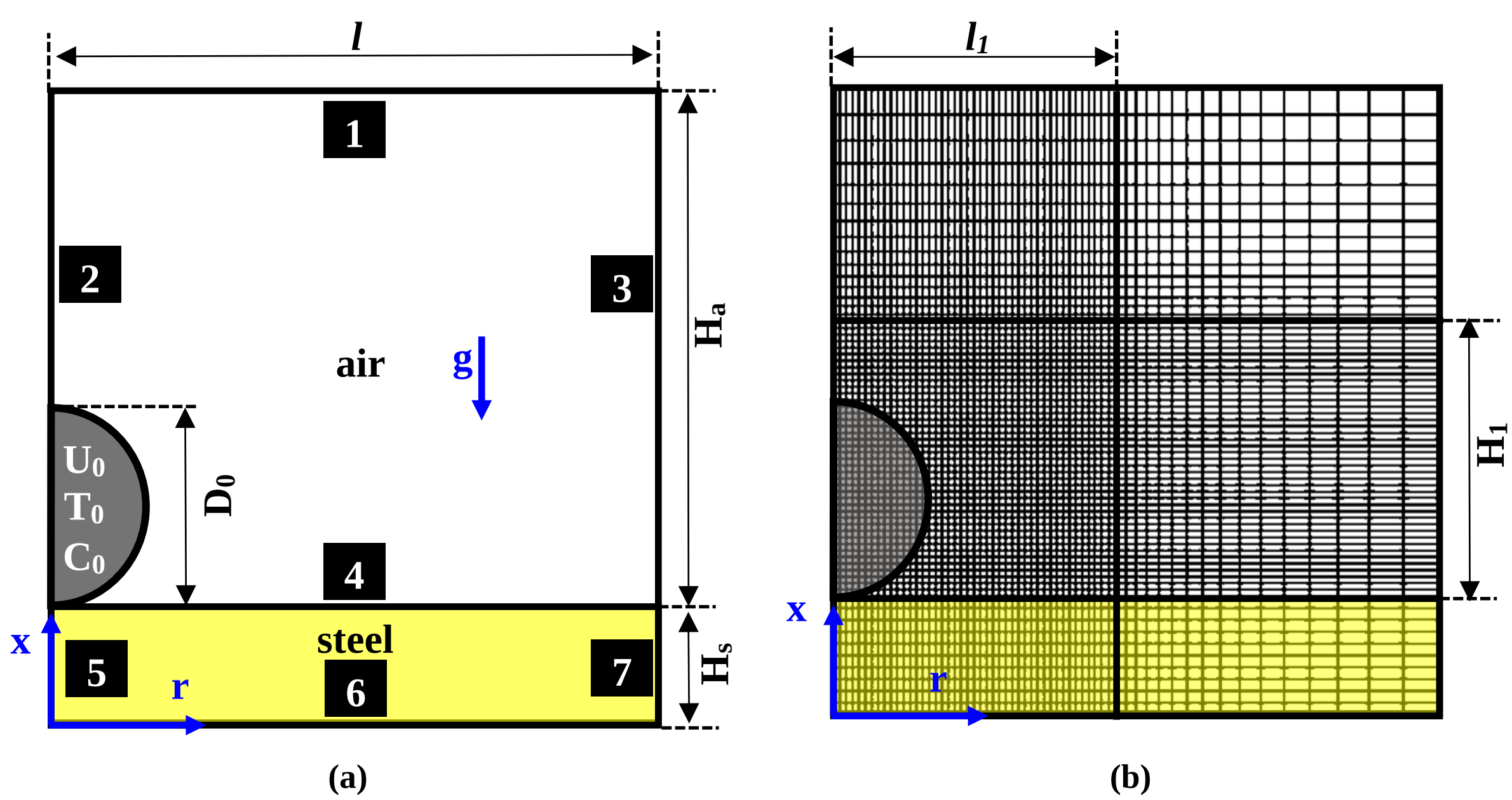

**Figure 1** (a) The computational domain: The boundaries are numbered 1-7. The substrate region is indicated by yellow color. (b) The computational mesh.

*2.2. Governing equations*

The current multi-physics process is a coupled multi-region, multi-phase, multi-component problem, involving multiphase fluid flow, surface wetting, heat transfer and solidification phase change, and species transport of metallic alloy components. We adopt a one-fluid formulation (Tryggvason *et al.* 2011) for the multi-phase region. To simplify the notation, the three-phase system is represented by denoting the aggregate solid- and liquid-alloy phases by subscript 1, $\alpha_1 = \alpha_{1,s} + \alpha_{1,l}$ where $\alpha$ is the volume fraction, while the air phase is denoted by subscript 2. Consequently, $\alpha_{1,s} + \alpha_{1,l} + \alpha_2 = \alpha_1 + \alpha_2 = 1$. In other words, subscript 1 refers to the entire alloy, regardless of its phase (solid or liquid). Therefore, a physical property of the (three-phase) mixture ($\zeta$) can be defined by the linear rule:

$$\zeta = \alpha_1 \zeta_1 + (1 - \alpha_1)\zeta_2. \tag{1}$$

The properties without these subscripts indicate the mixture properties. The alloy properties are computed based on a volume-weighted average,

$$\zeta_1 = \alpha_s \zeta_s + (1 - \alpha_s)\zeta_l, \tag{2}$$

for the density ($\zeta = \rho$) and thermal conductivity ($\zeta = k$), where subscripts $s$ and $l$ refer to the solidified and liquid (molten) alloy, or based on a mass-weighted average,

$$\zeta_1 = f_s \zeta_s + (1 - f_s)\zeta_l, \tag{3}$$

for the specific heat ($\zeta = c_p$) and mass diffusivity ($\zeta = D$), where $f_s$ is the alloy solid mass fraction, governed by the microsegregation model. Note that in our formulation, $\alpha_s$, $\alpha_l$, $f_s$, and $f_l$ are defined as a fraction of phase 1 (alloy) rather than the whole mixture, i.e.,

$$\alpha_s \equiv \frac{\alpha_{1,s}}{\alpha_1}, \alpha_l \equiv \frac{\alpha_{1,l}}{\alpha_1}, f_l \equiv \frac{\rho_l \alpha_l}{\rho_1}, f_s \equiv \frac{\rho_s \alpha_s}{\rho_1}.$$

Therefore, the following relations hold throughout the entire domain:

$$\alpha_1 + \alpha_2 = 1, \alpha_l + \alpha_s = 1, f_l + f_s = 1, \tag{4}$$

$$\rho_s \alpha_s = \rho_1 f_s = [\alpha_s \rho_s + (1 - \alpha_s)\rho_l] f_s = \left(\frac{f_s}{\rho_s} + \frac{1 - f_s}{\rho_l}\right)^{-1} f_s. \tag{5}$$

The problem entails three pairs of interfaces. The interface between the molten alloy and gas and the one between the solidified alloy and gas are considered sharp interfaces. These are captured here using the CLSVOF (Sussman & Puckett 2000, Son & Hur 2002) approach. On the other hand, the mushy region between the molten and solidified alloy is modelled by the EP method (Voller & Prakash 1987). The CLSVOF equations begin with the VOF part which gives an approximation, $\tilde{\alpha}_1$, to the alloy volume fraction by

$$\frac{\partial}{\partial t}(\tilde{\alpha}_1\rho_1) + \boldsymbol{\nabla}.(\tilde{\alpha}_1\rho_1\boldsymbol{u}) = 0, \tag{6}$$

where $\boldsymbol{u}$ is the (mixture) velocity, and bold symbols denote the vector or tensor quantities. Given the approximation $\tilde{\alpha}_1$, the level-set function, $\psi$, is initialized by (at each time step) (Albadawi *et al.* 2013)

$$\psi_0 = 0.75\Delta(2\tilde{\alpha}_1 - 1), \tag{7}$$

where $\Delta$ is the cell size (in the uniform block, see Figure 1b). Then, a re-initialization equation is solved to maintain the distance property of the level-set function:

$$\frac{\partial\psi}{\partial\tau} = \mathrm{sign}(\psi_0)(1 - |\boldsymbol{\nabla}\psi|), \tag{8}$$

where $\mathrm{sign}(.)$ is the sign function, and $\tau$ is an artificial time. Eq. (8) is solved until $\tau = 0.1\Delta$ to complete the re-initialization. Finally, the volume fraction is obtained by an approximate Heaviside function:

$$\alpha_1 = H(\psi) = \begin{cases} 0 & ;\psi < -\epsilon \\ \frac{\psi+\epsilon}{2\epsilon} + \frac{\sin\left(\frac{\pi\psi}{\epsilon}\right)}{2\pi} & ;|\psi| \le \epsilon, \\ 1 & ;\psi > \epsilon \end{cases} \tag{9}$$

where the interface thickness parameter is chosen as $\epsilon = 1.5\Delta$.

The continuity and momentum equations for an unsteady, incompressible (Boussinesq approximation for buoyancy), three-phase flow are

$$\frac{\partial\rho}{\partial t} + \boldsymbol{\nabla}.(\rho\boldsymbol{v}) = 0, \tag{10}$$

$$\frac{\partial}{\partial t}(\rho\boldsymbol{u}) + \boldsymbol{\nabla}\cdot(\rho\boldsymbol{u}\boldsymbol{u}) = \boldsymbol{\nabla}\cdot[\mu(\boldsymbol{\nabla}\boldsymbol{u} + (\boldsymbol{\nabla}\boldsymbol{u})^T)] - \boldsymbol{\nabla}p + \rho_B\boldsymbol{g} + \boldsymbol{S}_{m,u} + \boldsymbol{F}_\sigma, \tag{11}$$

where $p$ is the static pressure, $\boldsymbol{g}$ the gravitational acceleration, and $\mu$ the viscosity. $\rho_B$ in the Buoyancy term is computed by

$$\rho_B = \alpha_1\rho_l\left[1 - \beta_{T,l}(T - T_0) - \beta_{C,l}(C - C_0)\right] + \alpha_2\rho_2, \tag{12}$$

where $\beta_{T,l}$ and $\beta_{C,l}$ are the liquid thermal and solutal expansion coefficients, respectively, $T_0$ is a reference temperature, and $C_0$ the nominal solute concentration. The mixture temperature, $T$, and the solute concentration, $C$, are governed by the heat and mass transfer transport equations. In Eq. (11), $\boldsymbol{S}_{m,u}$ is the pressure drop within the mushy zone, which has two components: the flow suppression term and the relative advection term (Asai & Muchi 1978, Bennon & Incropera 1987a). This term can be written as

$$\boldsymbol{S}_{m,u} = \alpha_1 \left\{ -\frac{\mu_l}{K}\frac{\rho_1}{\rho_l}\boldsymbol{u} - \boldsymbol{\nabla} \cdot \left[ \rho_1 \frac{f_{s,B}}{(1-f_{s,B})} \boldsymbol{uu} \right] \right\}; \; f_{s,B} = \min(f_s, 0.99), \tag{13}$$

where $f_{s,B}$ is used to prevent division by zero, and $K$ is the permeability of the mushy zone, modeled by the standard Blake-Kozeny-Carman (BKC) model as (Asai & Muchi 1978, Bennon & Incropera 1987a)

$$K^{-1} = K_0^{-1} \left[ \frac{f_{s,B}^2}{(1-f_{s,B})^3} \right]; \; K_0 = \frac{\lambda_2^2}{180}. \tag{14}$$

In Eq. (14), $\lambda_2$ is the secondary dendrite arm spacing.

In Eq. (11), the surface tension (per unit volume), $\boldsymbol{F}_\sigma$, is calculated using the CLSVOF variant of the Continuum Surface Force (CSF) model as follows:

$$\boldsymbol{F}_\sigma = \sigma \kappa \delta(\psi) \boldsymbol{\nabla} \psi, \tag{15}$$

Here, $\sigma$ stands for the gas-alloy surface tension coefficient, $\kappa$ the local radius of curvature at the gas-alloy interface:

$$\kappa = \boldsymbol{\nabla} . \boldsymbol{n}, \tag{16}$$

$\boldsymbol{n}$ the unit normal vector at the interface:

$$\boldsymbol{n} = \frac{\boldsymbol{\nabla}\psi}{|\boldsymbol{\nabla}\psi|}, \tag{17}$$

and $\delta(\psi)$ the delta function approximated by

$$\delta(\psi) = \begin{cases} 0 & ; |\psi| > \epsilon \\ \dfrac{1 + \cos\left(\frac{\pi\psi}{\epsilon}\right)}{2\epsilon} & ; |\psi| \le \epsilon \end{cases}. \tag{18}$$

The surface wetting is accounted for by modifying the unit normal vector, $\boldsymbol{n}$, used in Eq. (17), within the cells adjacent to a solid wall by (Rettenmaier 2019)

$$\boldsymbol{n}_c = a\boldsymbol{n}_0 + b\boldsymbol{n}_w, \tag{19}$$

where $\boldsymbol{n}_c$ is the corrected value of the interface normal at wall-adjacent cells, $\boldsymbol{n}_0$ is the value computed by Eq. (17), and $\boldsymbol{n}_w$ the local wall unit normal. The coefficients $a$ and $b$ are given by

$$a = \frac{\cos\theta - \cos\theta_0 \cos(\theta_0 - \theta)}{1 - \cos^2\theta_0}, \tag{20}$$

$$b = \frac{\cos(\theta_0 - \theta) - \cos\theta_0 \cos\theta}{1 - \cos^2\theta_0}, \tag{21}$$

where $\cos\theta_0 = \boldsymbol{n}_0.\boldsymbol{n}_w$ and $\theta$ is the Contact Angle (CA). This angle can be estimated by the equilibrium CA, $\theta_E$, or by a DCA model accounting for CAH data. Here, when CAH data, i.e., the advancing and receding CAs ($\theta_A$ and $\theta_R$), is available, the Kistler DCA model (Kistler 1993, Roisman *et al.* 2008) is adopted, where $\theta$ is computed as a function of $\theta_E$, $\theta_A$, $\theta_R$, and the CL velocity, $V_{CL}$, which is estimated by a conventional model based on the fluid velocity at wall adjacent cells. The details of formulation of the Kistler DCA model and the conventional CL velocity estimation have been given in previous publications (Amani & Abdi-Sij 2024).

The heat transfer and phase change model determines the flow field temperature, $T$, by solving for the following energy equation in the multiphase region (see Figure 1), based on the EP method (Bennon & Incropera 1987a, b, Richter *et al.* 2017, Nandana & Janoske 2018) extended here to a three-phase problem as

$$\frac{\partial}{\partial t}\left(\rho c_p T\right) + \boldsymbol{\nabla} \cdot \left(\rho \boldsymbol{u} c_p T\right) = \boldsymbol{\nabla} \cdot (k\boldsymbol{\nabla}T) + \frac{\partial}{\partial t}\left(\alpha_1 \rho_1 L_f f_s\right), \tag{22}$$

where $L_f$ is the latent heat of solidification. Within the substrate region (Figure 1), a pure conduction equation is solved as

$$\boldsymbol{\nabla}.\left(k_w \boldsymbol{\nabla} T_w\right) = \rho_w c_{p,w} \frac{\partial T_w}{\partial t}, \tag{23}$$

where the subscript $w$ indicates the wall (substrate) properties.

To account for the mass transfer of the multi-component alloy, the following species transport equation is solved throughout the multiphase flow domain, based on the model by Bennon and Incropera (1987a) extended here to a three-phase problem as

$$\frac{\partial(\alpha_1\rho_1 C_1)}{\partial t} + \boldsymbol{\nabla}\cdot(\alpha_1\rho_1\boldsymbol{u}C_1) = \boldsymbol{\nabla}.(\alpha_1\rho_1 D_1\boldsymbol{\nabla}C_1) + \boldsymbol{\nabla}\cdot[\alpha_1\rho_1 D_1\boldsymbol{\nabla}(C_l - C_1)] - \boldsymbol{\nabla}\cdot[\alpha_1\rho_1\boldsymbol{u}(C_l - C_1)], \tag{24}$$

where $D_1$ is the solute diffusivity in the alloy phase (1), and $C_1$ is the solute mass fraction within the alloy in either liquid or solid state, i.e.,

$$f_s C_s + (1 - f_s)C_l = C_1. \tag{25}$$

Note that, in contrast to the momentum and energy equations, it is more convenient to write the species transport equation, Eq. (24), in terms of $C_1$ rather than the mixture property $C$, since for the current problem $C_2 = 0$. The field $C$ is computed only at the post-processing stage by $C = \alpha_1 C_1 + (1 - \alpha_1)C_2 = \alpha_1 C_1$. This automatically filters out the value of $C_1$ outside the liquid phase (phase 1) which has no physical meaning. Notably, $C$ represents the species mass fraction, which is a dimensionless variable. However, to comply with the majority of the solidification literature, we use the term "concentration" in place of mass fraction in the current text.

To close the preceding equations, the phase diagram data and microsegregation model should be added to the list of equations. Assuming a hypoeutectic system with a linear liquidus and solidus lines on the phase diagram, the liquidus ($T_{\text{liq}}$) and solidus ($T_{\text{sol}}$) temperatures are given by the following relationships:

$$T_{\text{liq}} = \begin{cases} T_{\text{melt}} + m_{\text{liq}} C_1 & ; C_1 \le C_{\text{eut}} \\ T_{\text{eut}} & ; C_1 > C_{\text{eut}} \end{cases}, \tag{26}$$

$$T_{\text{sol}} = \begin{cases} T_{\text{melt}} + m_{\text{liq}} C_1 / K_p & ; C_1 \le K_p C_{\text{eut}} \\ T_{\text{eut}} & ; C_1 > K_p C_{\text{eut}} \end{cases}. \tag{27}$$

In these expressions, $T_{\text{melt}}$ is the melting temperature of the pure solvent, $m_{\text{liq}}$ is the liquidus line slope on the phase diagram, $K_p$ is the partition coefficient, and $T_{\text{eut}}$ and $C_{\text{eut}}$ are the eutectic point temperature and concentration, respectively. In addition, assuming thermodynamic equilibrium at solidification front, implying infinitely fast solute diffusion in both phases (Bennon & Incropera 1987a), the conventional Lever rule can serve as the microsegregation model (Stefanescu 2015):

$$f_s = \frac{1}{1 - K_p}\left[\frac{T - T_{\text{liq}}}{T - T_{\text{melt}}}\right], \tag{28}$$

$$C_l = C_1\left[1 + f_s\left(K_p - 1\right)\right]^{-1}, \tag{29}$$

$$C_s = K_p C_l. \tag{30}$$

Alternatively, the Scheil rule can be adopted by replacing Eqs. (28)–(30) with (Stefanescu 2015)

$$f_s = 1 - \left(\frac{T_{\text{melt}} - T}{T_{\text{melt}} - T_{\text{liq}}}\right)^{\left(\frac{1}{K_p - 1}\right)} \tag{31}$$

$$C_l = C_1[1 - f_s]^{(K_p - 1)} \tag{32}$$

$$C_s = \frac{C_1 - (1 - f_s)C_l}{f_s + 0.001} \tag{33}$$

The effect of microsegregation model choices is explored in the results and discussion section.

Finally, it should be noted that we adopted a simplified binary alloy assumption. In multicomponent systems, interactions between solutes alter liquidus/solidus evolution and can shift segregation intensity or even modify local segregation patterns, which simplified binary models may underpredict or oversmooth (Du *et al.* 2007, Tu *et al.* 2016, Liu *et al.* 2026). Multicomponent models can provide higher fidelity by resolving these interactions. In addition, our computational geometry (Figure 1) and problem definition assumes a 2D axisymmetric splat morphology, which is valid for low-Weber number droplet deposition applications. At higher Weber numbers, 3D phenomena, like fingering or splashing, can occur which are beyond the scope of the present work.

### *2.3.A zero-concentration-flux condition across interface*

Since in the current one-fluid formulation, Eq. (24) is solved within the whole multiphase flow domain, the numerical diffusion error leads to an unphysical mass transfer across the droplet-gas interface, violating the solute mass conservation within the droplet. This error is accumulated over time and generally cannot be neglected. To address this issue, the idea is to introduce a solute mass flux at the droplet interface which is equal to the numerical diffusion flux in the opposite direction. In other words, due to numerical errors, the left-hand side of Eq. (34) is not zero and is equal to a numerical error term, say $\dot{m}_e$,

$$\frac{\partial}{\partial t}(\rho_1 \alpha_1) + \boldsymbol{\nabla} \cdot (\rho_1 \alpha_1 \boldsymbol{u}) = \dot{m}_e. \tag{34}$$

If a new corrected velocity field, $\boldsymbol{u}^*$, is defined such that

$$\rho_1 \alpha_1 \boldsymbol{u}^* = \rho_1 \alpha_1 \boldsymbol{u} + \rho_1 \alpha_1 \boldsymbol{u}_e, \tag{35}$$

where the extra velocity field, $\boldsymbol{u}_e$, satisfies the following relation:

$$\boldsymbol{\nabla} \cdot (\rho_1 \alpha_1 \boldsymbol{u}_e) \equiv -\dot{m}_e, \tag{36}$$

then, substituting Eqs. (35) and (36) into Eq. (34) yields

$$\frac{\partial}{\partial t}(\rho_1 \alpha_1) + \boldsymbol{\nabla} \cdot (\rho_1 \alpha_1 \boldsymbol{u}^*) = 0. \tag{37}$$

It means $\boldsymbol{u}^*$ is a velocity field with canceled-out mass flux across the interface. The Helmholtz theorem states that a vector field (with decaying condition at infinity) is uniquely determined by its divergence and curl. Based on this theorem, the vector field $\rho_1 \alpha_1 \boldsymbol{u}_e$ can be decomposed as the sum of a scalar field potential $\phi$ and a vector field potential $\boldsymbol{A}$: $\rho_1 \alpha_1 \boldsymbol{u}_e = \boldsymbol{\nabla}\phi + \boldsymbol{\nabla} \times \boldsymbol{A}$. Since it is only required that the divergence of $\rho_1 \alpha_1 \boldsymbol{u}_e$ satisfies Eq. (36), its rotational part is arbitrarily chosen to be zero ($\boldsymbol{\nabla} \times \boldsymbol{A} \equiv 0$). Therefore, this relation is simplified to

$$\rho_1 \alpha_1 \boldsymbol{u}_e = \boldsymbol{\nabla}\phi. \tag{38}$$

Substituting Eqs. (34) and (38) into Eq. (36) leads to

$$\nabla^2 \phi = -\left[\frac{\partial(\rho_1 \alpha_1)}{\partial t} + \boldsymbol{\nabla} \cdot (\rho_1 \alpha_1 \boldsymbol{u})\right]. \tag{39}$$

To summarize the algorithm, first, the Poisson's equation, Eq. (39), is solved for the scalar field $\phi$. Then, the correction flux $\rho_1 \alpha_1 \boldsymbol{u}_e$ is calculated by Eq. (38). Finally, the corrected flux $\rho_1 \alpha_1 \boldsymbol{u}^*$ is computed by Eq. (35) and is used in place of the original flux $\rho_1 \alpha_1 \boldsymbol{u}$ in Eq. (24). In other words, the final form of the species transport equation can be expressed as (in place of Eq. (24))

$$\frac{\partial(\alpha_1 \rho_1 C_1)}{\partial t} + \boldsymbol{\nabla} \cdot (\alpha_1 \rho_1 \boldsymbol{u}^* C_1) = \boldsymbol{\nabla}.(\alpha_1 \rho_1 D_1 \boldsymbol{\nabla} C_1) + \boldsymbol{\nabla} \cdot [\alpha_1 \rho_1 D_1 \boldsymbol{\nabla}(C_l - C_1)] - \boldsymbol{\nabla} \cdot [\alpha_1 \rho_1 \boldsymbol{u}^*(C_l - C_1)]. \tag{40}$$

*2.4. Non-dimensional parameters*

Since, this is the first use of the full hydrodynamics, wetting, heat transfer, phase change, and species transport for an impacting droplet, it is desirable to introduce all relevant dimensionless governing parameters. We keep the consistency with the parameters defined for the case without species transport, which has been covered in the literature (Attinger *et al.* 2000, Haferl *et al.* 2000). Performing a dimensional analysis using the governing equations described in section 2.2, all parameters are presented in Table 1. The specific values of parameters given in this table were calculated based on the nominal properties of the alloy AlA380 (see Appendix A), with $T_{\text{liq}}^0$ and $T_{\text{sol}}^0$ representing the liquidus and solidus temperatures at the nominal composition ($C_0$). Note also that these parameters are defined to be convenient for parametric studies, in such a manner to isolate the effect of each important dimensional property to a single dimensionless parameter. For instance, the variation of the superheat only changes SHP while keeping other dimensionless parameters constant. The Buoyancy Ratio (BR) is defined here as the ratio of the thermal to solutal buoyancy forces. Since the melt cools down (a temperature decrease) and enriches (a solute concentration increase) by approaching the solidification front, a negative value of BR indicates counteracting thermo-solutal buoyancy (acting in opposite directions), and vice versa.

**Table 1** The dimensionless governing parameters.

| Parameters | Acronym | Definition | Value |
|---|---|---|---|
| Weber number | We | $\frac{\rho_l V_0^2 D_0}{\sigma}$ | 84.73 |
| Ohnesorge number | Oh | $\frac{\mu_l}{\sqrt{\rho_l D_0 \sigma}}$ | $3.52 \times 10^{-4}$ |
| Equilibrium contact angle | $\theta_E$ | – | 90 |
| Darcy number | Da | $\frac{K_0}{D_0^2}$ | $2.088 \times 10^{-6}$ |
| Grashof number | Gr | $\frac{g\beta_{T,l}(T_0 - T_{\text{sol}}^0)\rho_l^2 D_0^3}{\mu_l^2}$ | $2.243 \times 10^4$ |
| Buoyancy ratio | BR | $\frac{K_p}{(1-K_p)} \frac{\beta_{T,l}(T_{\text{sol}}^0 - T_0)}{\beta_{C,l} C_0}$ | $-10.4$ |
| Stefan number | Ste | $\frac{c_{p,l}(T_{\text{liq}}^0 - T_{w,0})}{L_f}$ | 1.04 |
| Superheat parameter | SHP | $\frac{T_0 - T_{\text{liq}}^0}{T_{\text{liq}}^0 - T_{w,0}}$ | 0.06 |
| Prandtl number | Pr | $\frac{c_{p,l}\mu_l}{k_l} = \frac{\mu_l}{\rho_l D_{T,l}}$ | 0.0165 |
| Schmidt number | Sc | $\frac{\mu_l}{\rho_l D_l}$ | 107.14 |
| Mushy temperature parameter | MTP | $\frac{T_{\text{liq}}^0 - T_{\text{sol}}^0}{T_{\text{liq}}^0 - T_{w,0}}$ | 0.067 |
| Density ratios | $\rho_{gl}, \rho_{sl}$ | $\frac{\rho_k}{\rho_l}; k = g, s$ | $(\rho_{gl}, \rho_{sl}) = (4.61 \times 10^{-4}, 1)$ |
| Viscosity ratio | $\mu_{gl}$ | $\frac{\mu_g}{\mu_l}$ | 0.016 |
| Thermal diffusivity ratios | $D_{T,gl}, D_{T,sl}, D_{T,wl}$ | $\frac{D_{T,k}}{D_{T,l}}; k = g, s, w$ | $(D_{T,gl}, D_{T,sl}, D_{T,wl}) = (0.79, 2.01, 0.19)$ |
| Mass diffusivity ratio | $D_{sl}$ | $\frac{D_s}{D_l}$ | $10^{-3}$ |
| Partition coefficient | $K_p$ | $K_p = \frac{C_s^*}{C_l^*}$ | 0.131 |

Some other well-known parameters, like Reynolds (Re), Capillary (Ca), Stokes (Sto), and (dimensionless) liquidus slope ($S_L$) based on the mushy temperature difference in the liquid depend on the ones introduced in Table 1, and can be computed (if necessary) by

$$\mathrm{Re} = \frac{\sqrt{\mathrm{We}}}{\mathrm{Oh}} = \frac{\mathrm{We}}{\mathrm{Ca}} = \eta_{gl}\mathrm{Sto} = \frac{\mathrm{Ha}^2}{N}, \tag{41}$$

$$S_L \equiv \frac{T_{\mathrm{liq}}^0 - T_{\mathrm{melt}}}{T_{\mathrm{liq}}^0 - T_{w,0}} = \frac{m_{\mathrm{liq}} C_0}{T_{\mathrm{liq}}^0 - T_{w,0}} = \mathrm{MTP}\ K_p. \tag{42}$$

The objective parameters of engineering interest can also be introduced in non-dimensional form, as defined in Table 2, where $t_{\mathrm{solid}}$ is the solidification time, $d(t)$ is the spreading diameter of the splat on the wall, $h(t)$ is the maximum height of the splat, $\Lambda_{\mathrm{b}}(t)$ is the volume of the entrapped air bubbles in the splat, and $S(t)$ and $S_{SD}(t)$ are the mean and standard deviation of the splat solidified-layer thickness. Note also that $\tau_s$ is the approximate solidification time scale, given by (Gao & Sonin 1994, Schiaffino & Sonin 1997, Attinger *et al.* 2000)

$$\tau_s = \frac{\rho_l L_f D_0^2}{k_l\left(T_{\mathrm{liq}}^0 - T_{w,0}\right)}. \tag{43}$$

Of particular interest here are the maximum SF, $\mathrm{SF}_{\mathrm{max}}$, final SF, $\mathrm{SF}_\infty$, final AH, $\mathrm{AH}_\infty$, final MSh, $\mathrm{MSh}_\infty$, final maximum and minimum SDs, $\mathrm{SD}_{\infty,\mathrm{max}}$ and $\mathrm{SD}_{\infty,\mathrm{min}}$, and final ST and SN, $\mathrm{ST}_\infty$ and $\mathrm{SN}_\infty$.

**Table 2** The dimensionless objective parameters.

| Number | Name | Acronym | Definition |
|---|---|---|---|
| 1 | Dimensionless solidification time | $t_s^*$ | $\frac{t_{\mathrm{solid}}}{\tau_s}$ |
| 2 | Spread factor | SF | $\frac{d(t)}{D_0}$ |
| 3 | Aspect height | AH | $\frac{h(t)}{D_0}$ |
| 4 | Mean solidified-layer thickness | ST | $\frac{S(t)}{D_0}$ |
| 5 | Solidified-layer non-uniformity | SN | $\frac{S_{\mathrm{SD}}(t)}{D_0}$ |
| 6 | Macro-Shrinkage | MSh | $\frac{\Lambda_{\mathrm{b}}(t)}{\frac{\pi}{6}D_0^3}\times 100\%$ |
| 7 | Segregation degree | SD | $\frac{C(\boldsymbol{x},t) - C_0}{C_0}\times 100\%$ |

## 3. Numerical method

A new solver is developed for the solution of the governing equations based on the finite volume method Computational Fluid Dynamics (CFD) package OpenFOAM v8 (www.openfoam.org) and the "solidificationFoam" solver (Coleman 2020). The new solver structure is detailed in the supplementary material. All details of the discretization schemes, linear equation solvers, convergence tolerance, along with an OpenFOAM case file are provided in the supplementary material. A short summary of the most important settings is: Pressure–velocity coupling is handled using a PIMPLE algorithm with the momentum predictor step enabled. Thermophysical coupling within the solidification loop is controlled through a minimum of three and a maximum of 20 correction iterations, with a convergence tolerance of $10^{-4}$ to account for the strong coupling between thermal and composition fields during solidification. For the solution of the VOF equation, Eq. (6), the OpenFOAM's built-in approach, MPLICU (velocity-weighted Multidimensional Piecewise Linear Interface Calculation) is employed, which is a variant to Youngs' PLIC geometric reconstruction scheme (Youngs 1982, Tryggvason *et al.* 2011). Normalized residual tolerances of $10^{-6}$ for temperature in the substrate region, $10^{-10}$ for the energy and species transport in the multiphase region, and $10^{-8}$ for all other variables are adopted. All governing equations in both fluid and solid regions are discretized in time using a first-order implicit Euler scheme (Moukalled *et al.* 2016) with a CFL number of 0.2 for the volume fraction and 0.8 for the other variables. The main momentum advection term is discretized using a bounded second-order Gauss linear scheme (Moukalled *et al.* 2016), while additional nonlinear advective terms associated with relative motion and phase interaction forces are discretized using a pure second-order Gauss linear scheme (Moukalled *et al.* 2016). For the temperature and concentration equations, the primary advective fluxes are discretized using a first-order upwind scheme, while additional advection terms, associated with phase change, latent heat transport, and relative motion between phases, are discretized using the second-order Gauss linear formulation.

The boundary conditions and mesh layout are depicted in Figure 1b. An entrainment condition is applied on boundaries 1 and 3, whereas a no-slip velocity condition is imposed on boundary 4. Boundaries 2 and 5 are treated as the symmetry axis. From a thermal standpoint, boundary 4 is assigned a temperature-coupled condition, while boundaries 6 and 7 are maintained at a constant temperature of 473 $K$, which is the same as the substrate initial temperature. The domain is filled with stagnant air gas at the initial uniform temperature. A single spherical molten alloy droplet is placed in the multiphase region, touching the top of the substrate, with a uniform initial axial velocity ($V_0$), temperature ($T_0$), and composition ($C_0$). The multiphase flow domain is divided into four blocks (Figure 1b). The main block, of size $l_1$ ($= 7\ mm$) and $H_1$ ($= 4.5\ mm$), contains the droplet during its impact and spread, and is discretized by a uniform mesh with a resolution of 90 Cells Per initial droplet Radius (CPR). This mesh resolution is selected following a grid-independence study, documented in Appendix B. In the other three blocks, the mesh is smoothly coarsened using a geometric expansion ratio of 1.2. The mesh within the solid substrate consists of 20 cells (of uniform height) along the axial (x) direction.

## 4. Results and discussion

The overall model described for alloy droplet impact and solidification involves different phenomena, including the hydrodynamics of a multiphase flow and wetting, heat transfer and phase change, and MS, which make model verification challenging. If the model is assessed only in its fully coupled form, the errors and uncertainties of different sub-models can misleadingly mask each other in a limited number of validation cases, hindering the generalization of the whole model. To avoid this pitfall, the validation strategy adopted here deliberately separates and interrogates the model components in a systematic manner. The solver is therefore examined through a sequence of canonical test cases, each involving only one or a few physical phenomena, thereby isolating the uncertainties associated with the underlying sub-models. The evaluation continues with benchmarks of increasing physical and numerical complexity, by gradually introducing additional sub-model couplings. The validation benchmarks and the procedure for the sub-model evaluation are summarized in Table 3. According to this table, the validation process is initiated in section 4.1, and, finally, the whole modeling framework is assessed in section 4.5. In section 4.5, in addition to analyzing the impact, solidification, and segregation degree in the droplet, the influence of species transport on an alloy droplet impact dynamics, which has been neglected in all previous studies, is scrutinized.

To highlight the limitations of the present validation strategy, it has to be mentioned that the steps in Table 3 do not include an explicit quantitative assessment of porosity or bubble formation (macro-shrinkage) against experiment, and the choice of the proper grid resolution to capture the related phenomena is solely based on grid-independence tests (Appendix B). In addition, a standard benchmark on casting in molds has been adopted in section 4.4, due to the lack of MS databases on solidifying-droplet impact. Developing reliable quantitative experimental benchmarks on porosity formation or MS for solidifying-droplet impact in the future is required for a more rigorous assessment of the model.

**Table 3** The specification of the validation benchmarks.

| Benchmark | Description | Section | Coupled sub-model(s) | Experimentally or analytically validated data |
|---|---|---|---|---|
| 1,2 | Pure hydrodynamics droplet-wall impact | 4.1 | Multiphase flow, wetting | Droplet shape evolution, spread factor, aspect height |
| 3 | Pure metal droplet impact and solidification | 4.2 | Multiphase flow, wetting, heat transfer, phase change | Spread factor, contact angle, solidified-layer thickness |
| 4,5 | Transport of a scalar confined within the liquid phase | 4.3 | Multiphase flow, zero-concentration-flux model | Concentration profiles |
| 6 | Alloy solidification inside a mold | 4.4 | Flow, heat transfer, phase change, mass transfer | MS map and profiles |
| 7 | Alloy droplet impact and solidification | 4.5 | All sub-models | Droplet shape evolution, spread factor |

### *4.1. Hydrodynamics of droplet-wall impact*

To assess the performance of the model over a representative range of relevant impact dynamics, of our interest, in the absence of heat and mass transfer, two experimentally documented water droplet-impact cases are selected. Together, these cases allow the model to be examined under both capillary- and inertia-dominated conditions. The first validation case corresponds to a low-Weber-number impact ($\mathrm{We} = 1.81$) and is based on the experiments by Roisman *et al.* (2008). The second case represents the moderate-Weber-number regime ($\mathrm{We} = 52$), drawn from the work by Rioboo *et al.* (2002). The physical and geometrical parameters of both benchmarks are summarized in the supplementary material. A grid resolution of 60 CPR is sufficient, based on a grid study. The geometrical parameters of the computational domain (Figure 1) for these cases are chosen as $l/D_0 = 3.5$, $H_a/D_0 = 5$, $l_1 = 5\ mm$ , and $H_1 = 4.5\ mm$.

The temporal variation of the spread factor and aspect height is shown in Figure 2. Around $t \approx 9\ ms$ for the first benchmark ($\mathrm{We} = 1.81$) and $t \approx 5\ ms$ for the second benchmark ($We\ =\ 52$), the spread factor reaches a peak and changes sign, indicating the end of the advancing spread and the onset of retraction, which is well captured by the present simulation. The relative errors of $\mathrm{SF}_{\max}$ prediction for the present model, with respect to the experimental measurement, are 3% and 1% for benchmarks 1 and 2, respectively, which are close to these values by the reference numerical simulations, i.e., 1.7% and 4%. On the other hand, $\mathrm{SF}_{\infty}$ for benchmark 1 shows 12% and 5% errors with the present and reference numerical methods, respectively. This difference can be mainly attributed to the different CL velocity computation algorithms used in the present and reference (Roisman *et al.* 2008) simulations. The recoil in the second test case ($We = 52$) is much larger, which eventually leads to a partial-rebound regime as shown in Figure 3. This figure illustrates droplet shape evolution for the second case. The present model successfully predicts the impact dynamics and deformation behavior, during both the spreading and final partial-rebound phases. According to these comparisons, the modeling strategy is reasonably capable of handling a range of impact dynamics and wettability.

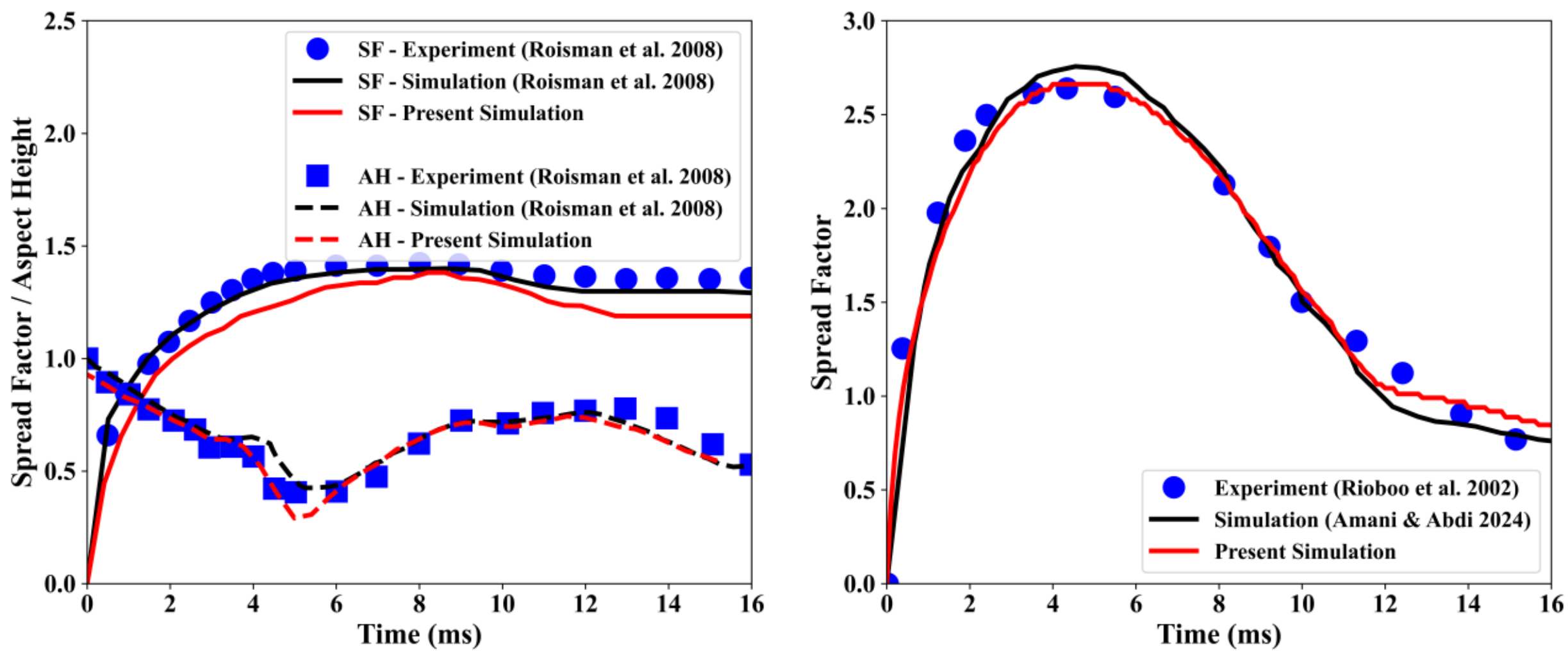

(a) (b)

**Figure 2** Temporal evolution of the spread factor and aspect height. a) Benchmark 1 (We = 1.81): comparison of the present numerical results against the experimental data and reference simulation (Roisman *et al.* 2008). b) Benchmark 2 (We = 52): the present results against experimental measurements (Rioboo *et al.* 2002) and previous numerical results (Amani & Abdi-Sij 2024).

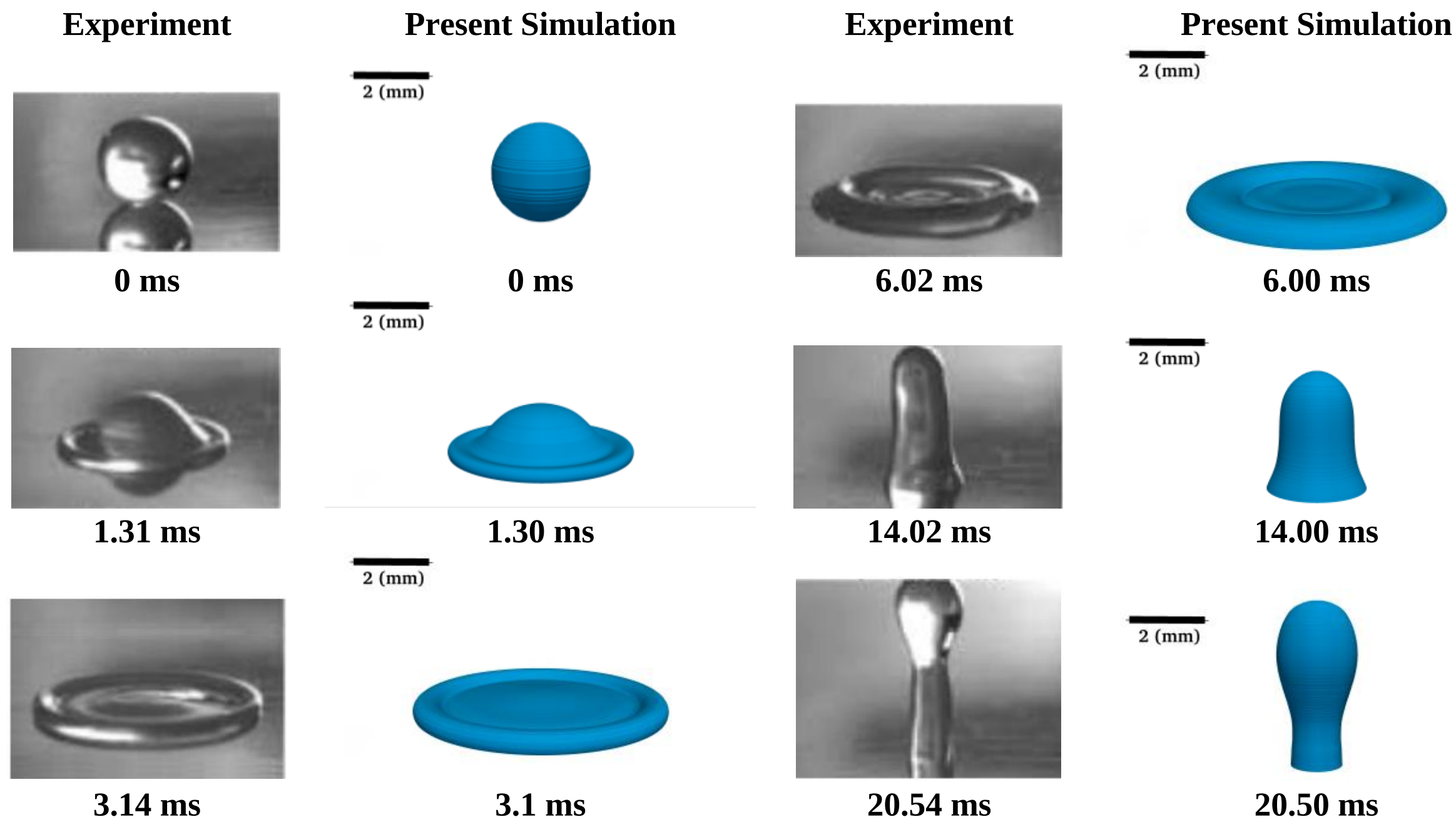

**Figure 3** Benchmark 2 (We = 52): The droplet shape evolution. The comparison of the present simulation with the experimental images (Rioboo *et al.* 2002).

*4.2. The solidification of a pure metal droplet impacting onto a solid wall*

To assess the accuracy of the heat transfer and phase change models, a dataset (benchmarks 3), provided by Pasandideh-Fard *et al.* (1998), involving a tin droplet impacting onto a steel surface, surrounded by Argon gas, is employed. Our analysis focuses on a single-region approach, where only the multiphase flow domain is considered, see Figure 1, and a cooling convective condition is applied to the solid substrate surface, as suggested by Pasandideh-Fard *et al.* (1998). The specification of the benchmark and its important dimensionless variables along with all relevant properties are summarized in the supplementary material. The schematic geometry and the computational grid are illustrated in Figure 1, with the specific choices of $l/D_0 = 3.5$, $H_a/D_0 = 5$, $l_1 = 5\,mm$, and $H_1 = 4.5\,mm$. A grid resolution of 60 CPR is used for the numerical solutions, based on a grid-independence test.

The temporal variations of the spread factor, solidified-layer thickness, and contact angle are plotted in Figure 4. The overall agreement of the present simulations with the experiments is satisfactory. However, there exist some discrepancies which should be analyzed further. The time and value of the peak spread factor are underpredicted by both the present modeling and the reference numerical simulation (Pasandideh-Fard *et al.* 1998), compared to the experiment. This deviation is primarily attributed to the fingering phenomenon, i.e., pronounced azimuthal rim instabilities, which appears between 2 to 5 ms and then fades away (see the experimental images provided by Pasandideh-Fard *et al.* (1998)). This phenomenon could not be accounted for in the present as well as the reference (Pasandideh-Fard *et al.* 1998) 2D-axisymmetric simulations and needs a computationally-intensive 3D simulation, like in (Amani & Abdi-Sij 2024), which is not the focus of the present work.

The variation of the dynamic contact angle, shown in Figure 4b, is well reproduced by the present simulation. The smooth transition between advancing and recoiling stage, advocated by gradual decrease in the dynamic contact angle, leads to a slow recoil phase and a small slope of spread factor reduction which is well captured by the simulation. Nevertheless, the value of the contact angle at the end of the recoiling stage is overpredicted, compared to the experimental data. This deviation can be connected to the uncertainties of the dynamic contact angle model over the last phase of the impact. The dimensionless solidified-layer thickness shown in Figure 4c demonstrates good agreement with the analytical approximation. This suggests that the 1D analytical model, which is commonly-used for developing semi-analytical correlations (Poirier & Poirier 1994), works fairly well at providing a reasonable estimate of this parameter. It should be noted that, in the second half of the solidification period till 6 ms when the solidification completes, the 1D assumption is not valid and the analytical model slightly underpredicts the solidified-layer thickness.

Despite the observed limitations, the global impact dynamics, wetting, heat transfer, and phase change are consistently represented by the model. To provide more evidence, these aspects are further evaluated in sections 4.4 and 4.5, in conjunction to the other phenomena.

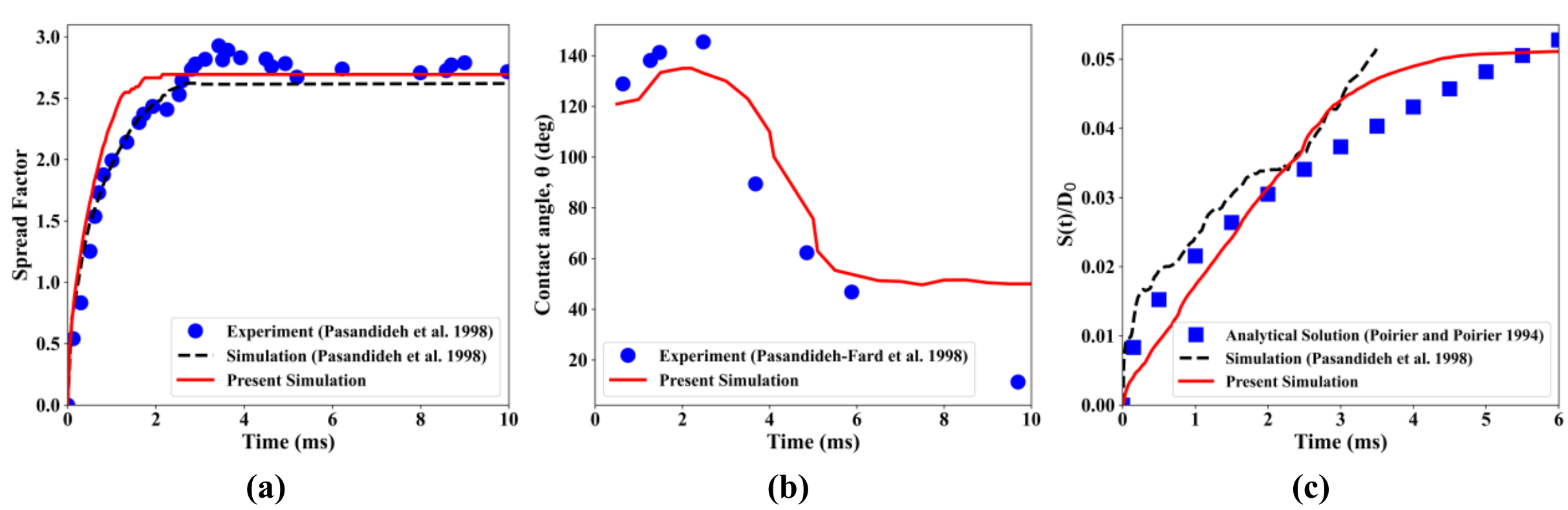

**Figure 4**: Benchmark 3 (Pasandideh-Fard *et al.* 1998): the spread factor (a), contact angle (b), and dimensionless solidified-layer thickness (c) versus time. Present simulations against the reference simulation and experimental measurements (Pasandideh-Fard *et al.* 1998) and 1D approximate analytical solution (Poirier & Poirier 1994).

### *4.3. Transport of a scalar field confined within the liquid phase*

The validation of the capability of the present model to impose a zero-concentration-flux at the droplet interface is critical for conserving the solute mass within the droplet and accurate prediction of the segregation degree. The benchmarks introduced by Jain and Mani (2023) for the same goal within the framework of diffuse-interface PF modeling are employed here. These benchmarks examine the transport of a passive scalar around a gas bubble in a 2D channel of height $H = 0.1\ m$. The computational domain, illustrated in Figure 5 (top left), consists of a square with periodic boundary conditions in the streamwise direction and no-slip walls on the Lower (L) and Upper (U) boundaries. A gas bubble of radius $R = 0.02\ m$ is initially positioned at the center of the domain and is surrounded by a liquid phase. The scalar transport is restricted to the liquid phase. Initially, the scalar concentration is zero throughout the domain. Dirichlet boundary conditions are imposed on the U and L walls, with concentrations $c_U = 1$ and $c_L = 0$, respectively. In the absence of a bubble, the steady-state concentration profile is linear, with a slope of $(c_U - c_L)/H = 10$. The presence of the bubble perturbs this profile, making this a demanding test for scalar-transport models.

The problem is investigated under two distinct conditions, corresponding to different bubble states, in order to pose different numerical challenges and assess model performance. These include: (i) a stationary bubble case with zero velocity throughout the domain, simulated on a $512 \times 512$ grid (benchmark 4); (ii) a moving, non-deformable bubble case, in which a uniform fluid velocity of $u = 1\ m/s$ is imposed everywhere in the domain, using a $128 \times 128$ grid (benchmark 5). Since the velocity field is uniform in the second case, the terminal concentration field in the frame attached to the moving bubble is steady and the same as the concentration field of the first case in the stationary frame. However, solving for the second case in a stationary frame with periodic boundary conditions poses more challenge to the solver, compared to the first case, due to the numerical errors of the fluxes. To maintain a uniform velocity field in benchmark 4 and benchmark 5, the momentum equation (Eq. (11)) is excluded from the solution loop and is therefore not solved during the simulation. The relevant physical properties are summarized in the supplementary material. For the sake of model assessment, a reference solution is obtained with a simple single-phase approach, which does not require interface capturing, let alone the interface concentration-flux correction. For this solution, the computational domain consists of the liquid-phase only, removing the circular gas bubble region in Figure 5 (top left), and applying a zero-gradient Neumann condition directly along the inner circular boundary of the liquid phase. To accurately capture the boundary-layer dynamics at the circular interface, an O-grid four-block boundary-fitted mesh topology is employed for this case, with a cell size comparable to the one used for the multi-phase simulations. For this goal, each of the four blocks in the O-grid is discretized by a $160 \times 160$ structured grid.

The steady-state scalar concentration fields obtained with the present CLSVOF model (with the proposed concentration-flux correction introduced in section 0) for benchmark 4 (stationary bubble) and benchmark 5 (moving non-deformable bubble) are compared with the reference single-phase solution and reference PF results in Figure 6. The results are extracted at $t = 2\ s$ when the steady-state condition prevails in both benchmarks, and the bubble is at the center of the domain in the moving bubble case (benchmark 5). For both benchmarks, the present CLSVOF simulations (Figure 6c and f) closely match the single-phase baseline solutions (Figure 6b and e). For a more quantitative comparison, the concentration profiles across a vertical cross-section, shown by the dashed line in Figure 5 (top left), are plotted in Figure 5. The present CLSVOF results obtained for both the stationary and moving bubble conditions coincide with each other and are in fine agreement with the single-phase reference solution, indicating that scalar transport is effectively restricted to the liquid region. To highlight the effect of the present concentration-flux correction algorithm, the result of the CLSVOF without flux correction, i.e., using Eq. (24) instead of Eq. (40), is shown in this figure. The larger deviation of this result from the reference solution is evident in this figure. Note that this profile is extracted at $t = 2\ s$ to be consistent with other results; and, in fact, the profile without flux correction does not reach a steady state condition, and the deviation with the reference solution grows as the solution proceeds further, due to the leakage of the scalar concentration from the interface. According to the comparison made in Figure 5, the PF results exhibit more noticeable deviations from the single-phase baseline, especially for the moving bubble case, since this case is more challenging due to the simultaneous action of advection, diffusion, and interface motion. The larger errors of the PF results, especially in the moving bubble case, may stem from a level of unphysical smearing observed in the shape of the bubble (Figure 6d) which is not fully consistent with the spherical shape of the bubble during pure translation. The reference paper did not mention any reason for this error in the shape of the bubble using PF method. This improved accuracy of the present CLSVOF approach highlights the superior capability of the current model to confine a scalar field and minimize artificial diffusion errors across the interface.

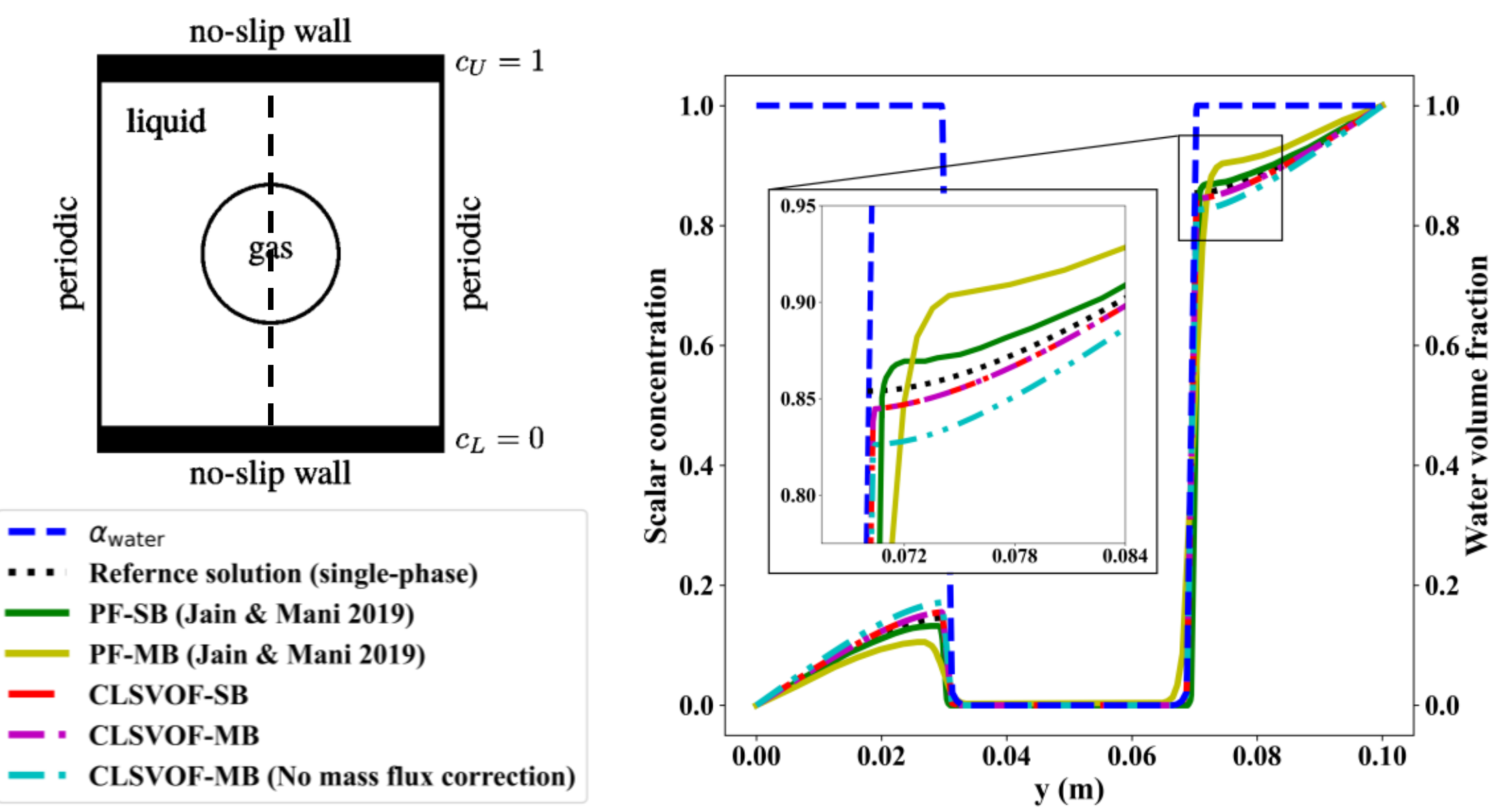

**Figure 5** Benchmark 4 and 5: the schematic computational domain (top left) (Jain & Mani 2023). Right) The liquid volume fraction ($\alpha_{\text{water}}$) and the steady-state scalar concentration ($C$) profiles over the vertical section, illustrated by the dashed line in part a. A comparison of the reference solution, present CLSVOF model predictions, and the PF model predictions (Jain & Mani 2023). SB and MB stand for the stationary bubble and moving non-deformable bubble cases, respectively.

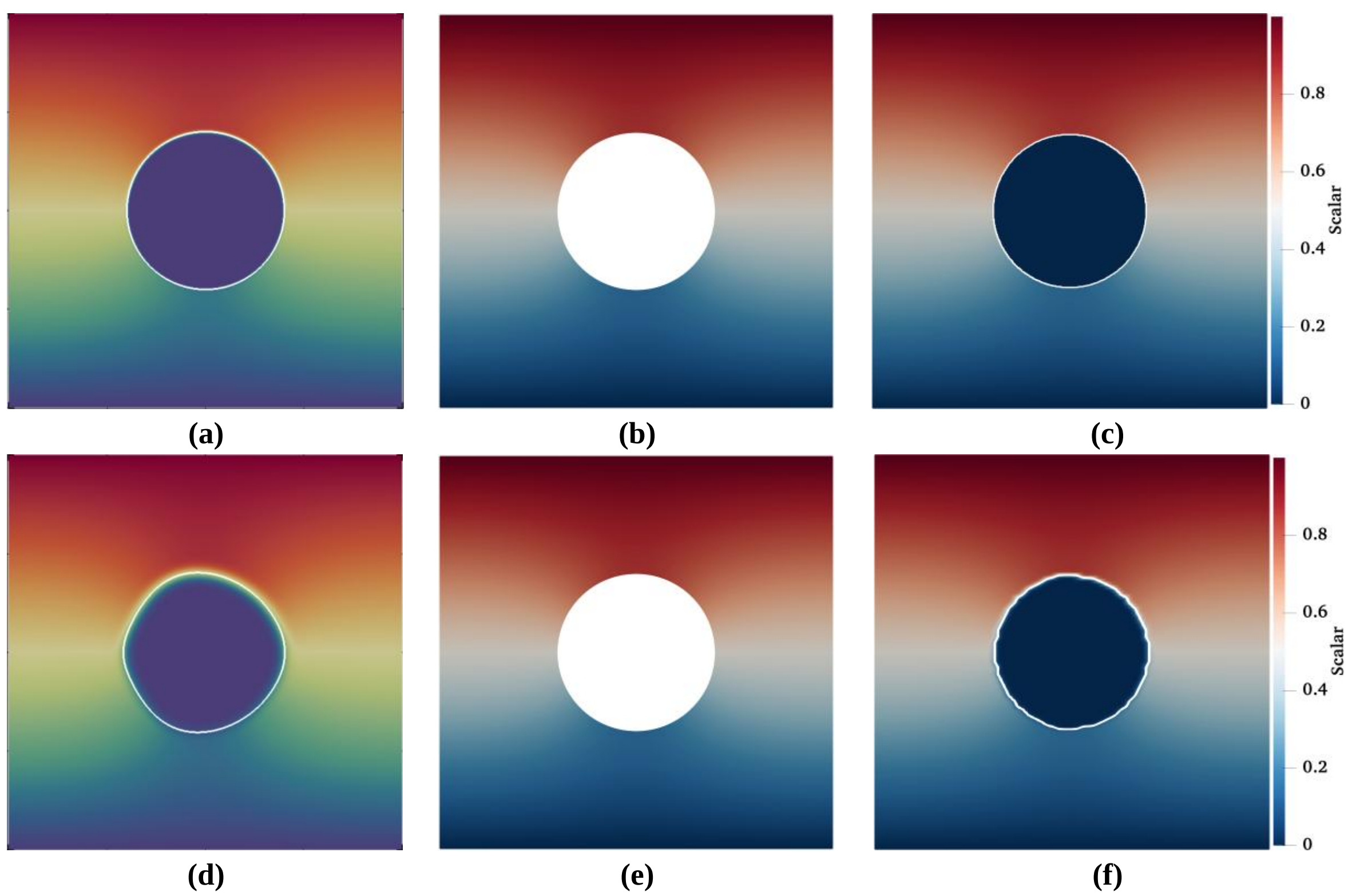

**Figure 6** The steady-state scalar concentration ($C$) distribution within the channel. Top row: stationary bubble (benchmark 4), and bottom row: moving bubble (benchmark 5). The PF model (Jain & Mani 2023) (a,d), the reference single-phase model (b,e), and the CLSVOF model (c,f).

### *4.4. MS in alloy solidification*

The segregation map measured for an alloy solidification inside a mold is a challenging benchmark to assess MS, governed by mass transfer (solute transport) coupled with heat transfer and fluid flow. Benchmark 6 considers the solidification of a lead-tin Pb-48%Sn alloy in a rectangular mold, based on the experimental study by Hebditch and Hunt (1974), a database that has been widely used for validating numerical models of alloy solidification. We designate two simulation setups: The first one is a single-region computational domain, similar to the one used in the previous numerical simulations (Chen & Shen 2018). In this setup, the mold is considered as a rectangular region, with a length of 100 mm and a height of 60 $mm$, completely filled with the alloy. The alloy region is shown by the red dashed box in Figure 7a. The molten alloy is initially quiescent, at its nominal composition $C_0$ $(= 0.48)$ and a uniform temperature $T_0$ $(= 489.15\ K)$. Cooling is applied through the left wall of the mold, while the remaining three walls (sides of the red dashed box) are assumed to be perfectly insulated. The thermal boundary condition at the cooled wall consists of a heat flux with a convective heat transfer coefficient of $300\ Wm^{-2}K^{-1}$ and temperature of $298.15\ K$. No-slip boundary condition and zero-gradient solute concentration are imposed on all walls of the mold. The thermophysical properties used in the simulation are summarized in the supplementary material.

The second simulation setup, shown in Figure 7a, is a two-region computational domain with a multiphase flow region, composed of the alloy and a 20 mm-high air layer above its top surface, and a 60 mm-thick solid-wall region on its left-hand side. This setup is designed here to assess the MS model along with the zero-concentration-flux boundary condition at the alloy-air interface and thermal coupling at the alloy-wall interface. The outer boundaries of the air and wall area are perfectly insulated. In addition, a very low thermal diffusion coefficient is chosen for air and solid-wall materials (to act as a thermal insulator) along with imposing a zero-velocity condition within whole air region throughout the simulation to suppress shear-induced movement at the air-melt interface. These are necessary for the two-region setup to be consistent with the boundary conditions of the original experiment and the single-region setup. A uniform grid with cubic cells of size $1\ mm \times 1\ mm$ is used within all regions which is consistent with the grid resolution used for the reference solution (Chen & Shen 2018).

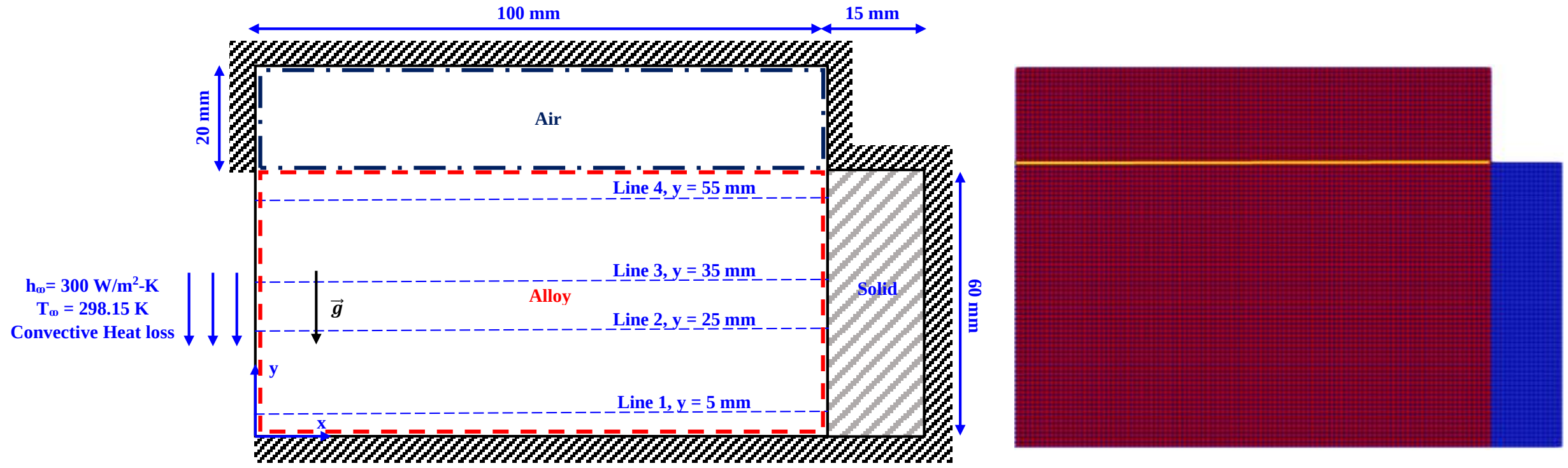

(a) (b)

**Figure 7** Benchmark 6: a) schematic computational domain, showing the alloy zone (red dashed box), the extra air zone (dark blue dashed-dotted box) at the top of the alloy, and the solid wall region (hatched area) at the right side of the alloy. b) The computational grid inside these regions. In the single-region setup, only the alloy region is considered. In the two-region simulation, a multiphase flow domain, including both the alloy and the top air region, is thermally coupled with the solid-wall region.

Figure 8 illustrates the evolution of temperature contours in the alloy domain, with superimposed melt velocity vectors and the mushy region borders indicated by the solid fraction iso-values, corresponding to the onset of solidification ($f_s = 0.05$) and complete solidification ($f_s = 0.95$), for both the two-region setup (left) and the single-region setup (right), at different time instances. At early times ($t = 150\ s$), the temperature contours are closely packed near the left cooled wall for both setups, indicating strong convective heat transfer and the rapid establishment of a solidification front. Upon $t = 150\ s$, a very thick mushy region is formed within the mold which occupies about half the mold volume. The melt flow penetrating this mushy region reaches the cold solidified alloy front ($f_s = 0.95$), which induces a downward thermal buoyancy force in the nearby melt. Note that the large Rayleigh number ($\mathrm{Ra} = \mathrm{Gr}\,\mathrm{Pr} = 1.837 \times 10^7$) for this case shows a strong natural convection. On the other hand, the solute rejection ahead of the solidification front within the mushy region modifies the liquid density field. Because lead is denser than tin, enrichment of tin in the liquid phase reduces the local density, which imposes an upward solutal buoyancy force on the melt. Therefore, the thermal and solutal buoyancy forces counteract each other in this problem which is evident from the sign of $\mathrm{BR} = -0.052 < 0$. Since the buoyancy ratio (BR) magnitude is much less than unity, the solutal buoyancy dominates and the melt pushes upward near the left wall. As a result, a clockwise natural convection is formed within the mold. All these complicated phenomena are captured by the both single- and two-region models, and their results are in fine agreement.

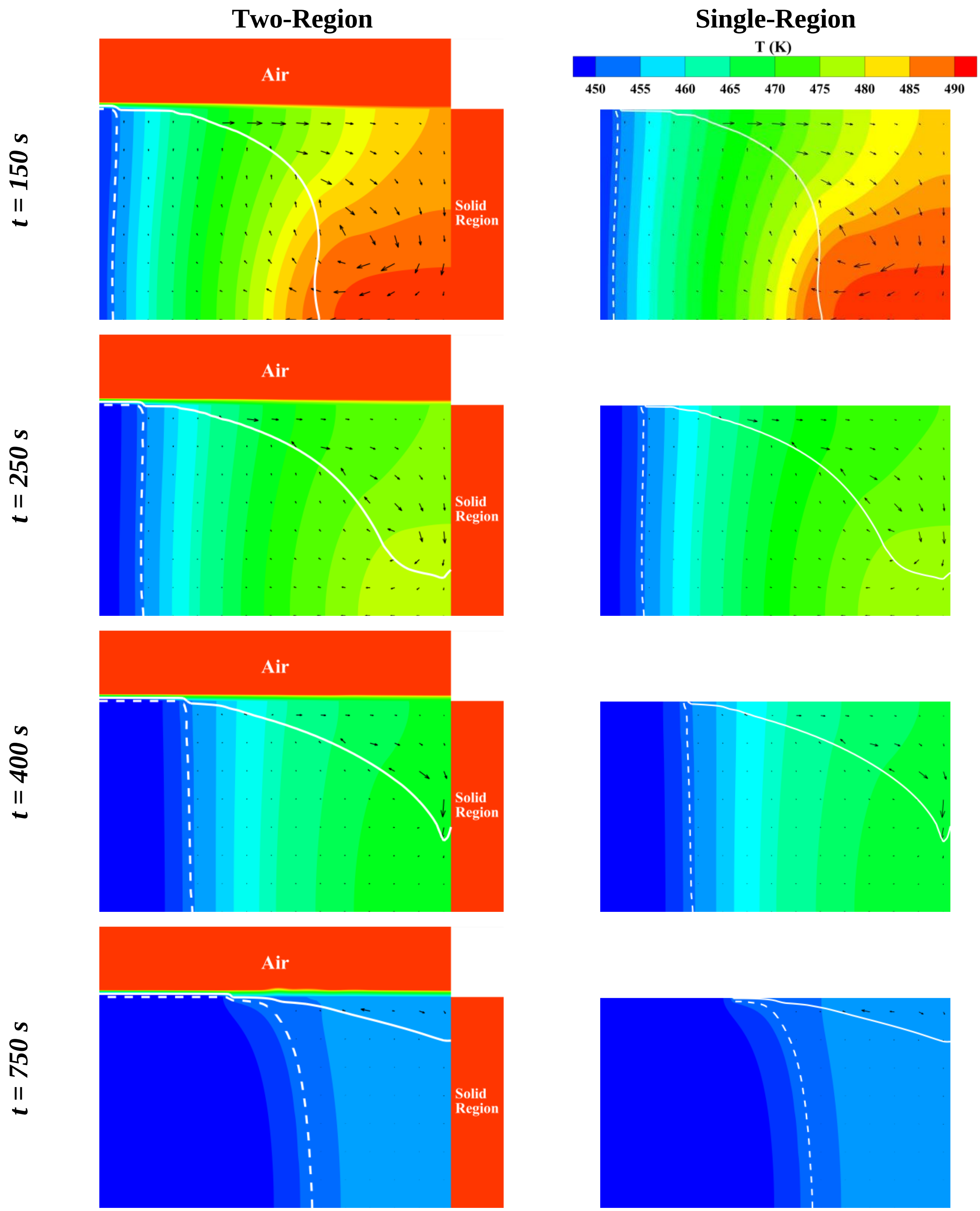

**Figure 8** Benchmark 6: The temperature contours with superimposed velocity field vectors and solid fraction iso-lines (the white dashed line indicates $f_s = 0.95$ and the white solid line $f_s = 0.05$) at different time instances. Left column: Two-region simulation, right column: single-region simulation.

Figure 9 compares the segregation map at $t = 400\ s$ predicted by the present single- and two-region models with the one reported by Chen and Shen (2018) (using a finite element discretization). Since the solidification commences at the left wall, negative SD regions are formed there due to the solute rejection to the melt. On the other hand, this rejected solute is accumulated in the melt, generating a solute-rich region and a large positive SD near the top-right portion of the mold, where the solidification completed according to Figure 8. As it can be observed in Figure 9, the two-region simulation produces a MS map consistent with those obtained using the single-region approach and the reference numerical simulation from the literature (Chen & Shen 2018).

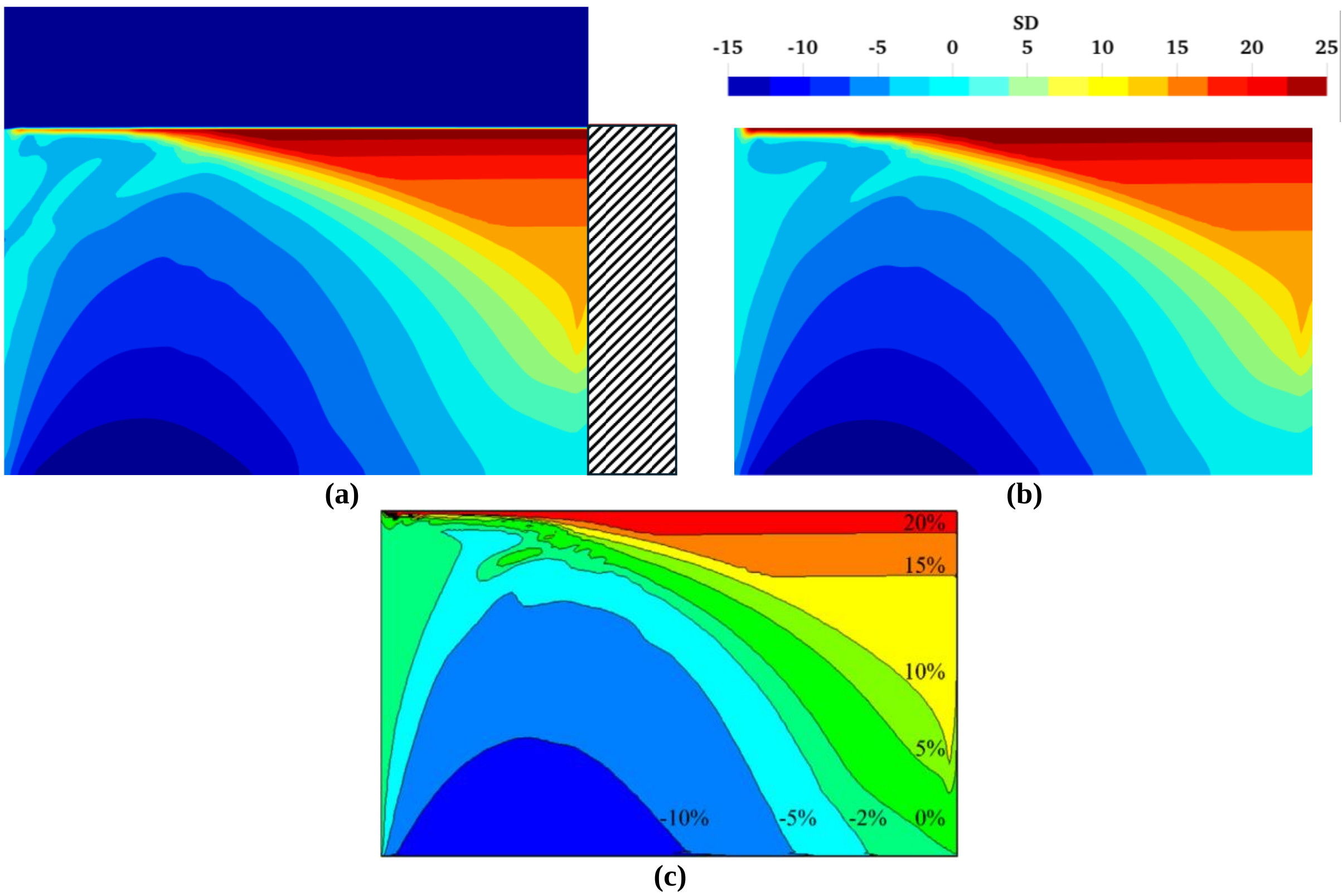

**Figure 9** Benchmark 6: the MS map (SD contours) at $t = 400\ s$. a) Two-region simulation, b) single-region simulation, and c) reference numerical simulation (Chen & Shen 2018).

To provide a more quantitative assessment and evaluation against experimental measurements, the final segregation degree (SD) profiles along horizontal lines at four different mold heights shown in Figure 7a (y = 5, 25, 35, and 55 mm) versus distance from the cooled wall are presented in Figure 10. At the lowest sampling height (y = 5 mm, figure 10a), all numerical solutions capture the strong negative segregation near the cooled wall, followed by a gradual recovery toward the nominal composition with increasing distance from the wall. This behavior is consistent with the early formation of a solidification front at the left boundary (Figure 8), where solute rejection enriches the adjacent liquid and drives solutal convection. The present two- and single-region simulations show very close agreement with each other and with the reference simulation, all in reasonable agreement with the measurements. At upper heights (y = 25 mm, y = 35 mm, and y = 55 mm, figure 10b-d), the SD profiles reveal a transition from negative to positive segregation across the mold width with a larger peak SD as sampling height elevates. This trend is consistent with the direction of the solidification front movement in Figure 8 and the accumulation of the tin-rich, lighter melt, near the upper-right part of the mold. Both the present single- and two-region SD profiles closely match the reference numerical results. Minor discrepancies between the experimental and numerical data persist, particularly in regions of weak segregation gradients, but considering the complexity of the problem, the overall agreement between the model and experiment is satisfactory.

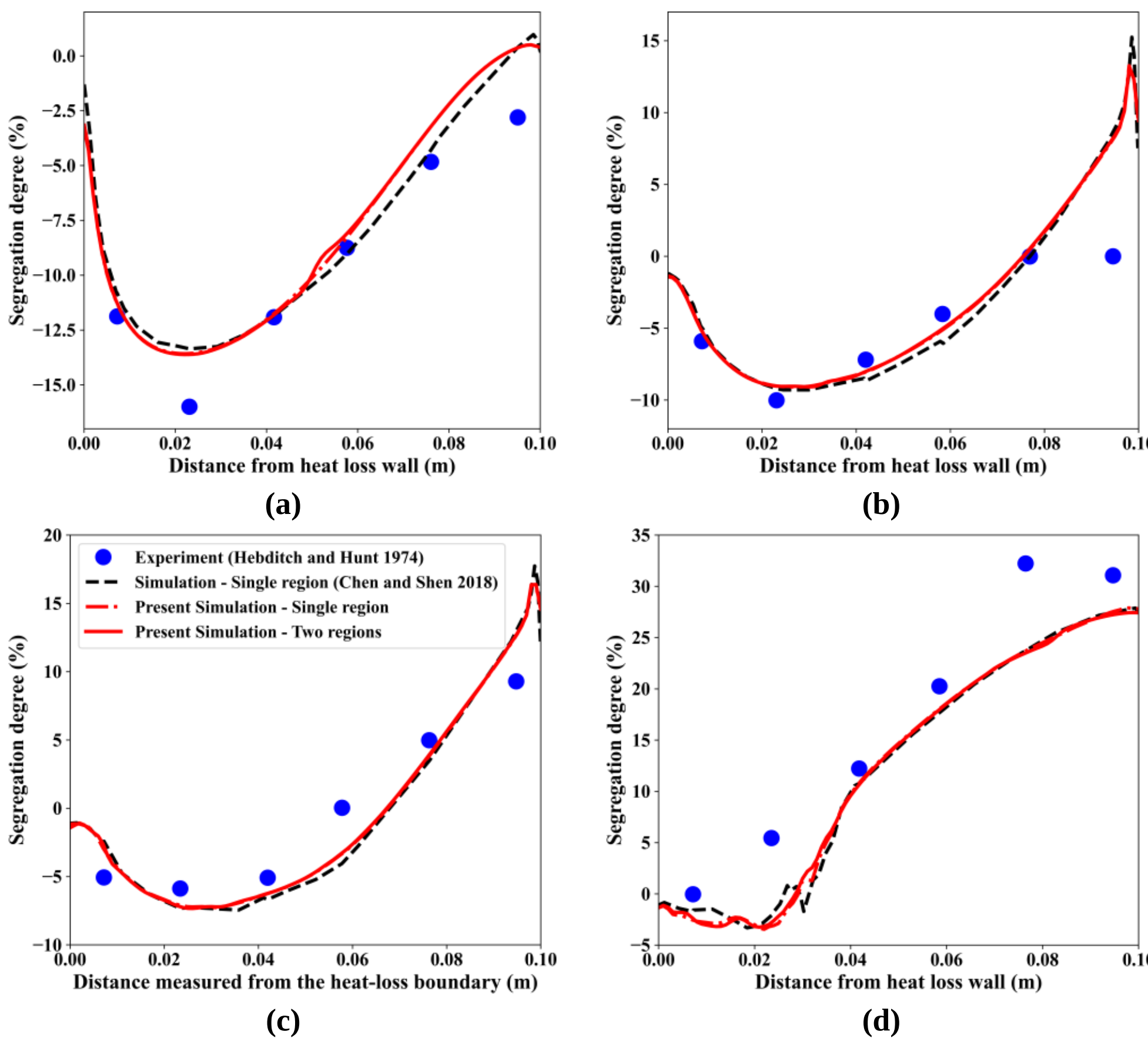

**Figure 10** Benchmark 6: the final SD profiles at different mold heights (Figure 7a): a) y = 5 mm, b) y = 25 mm, c) y = 35 mm, and d) y = 55 mm. Comparisons of the present single-region and two-region simulations against the experimental measurements (Hebditch & Hunt 1974) and reference numerical simulation (Chen & Shen 2018).

To assess the performance of the solute-concentration-flux correction in the present problem, Figure 11 illustrates the variation of the total solute mass in the alloy phase (both solid and liquid phases) within the mold predicted by the two-region simulations under two scenarios: (i) the default model whose results are discussed in Figure 8-Figure 10, and (ii) the model omitting the mass flux correction. Note that the total solute mass is computed by the volume integral $\int \alpha_1 \rho_1 C_1 \, dV = \int \rho_1 C \, dV$ over the whole multiphase flow domain. In the case with no mass flux correction, the total solute mass exhibits a continuous decline over time as the simulation continues due to the numerical diffusion error at the alloy-air interface. The level of solute mass loss depends on the process duration. In contrast, the present concentration-flux correction algorithm effectively conserves the solute mass throughout the simulation time, which underscores the critical importance of the flux correction step.

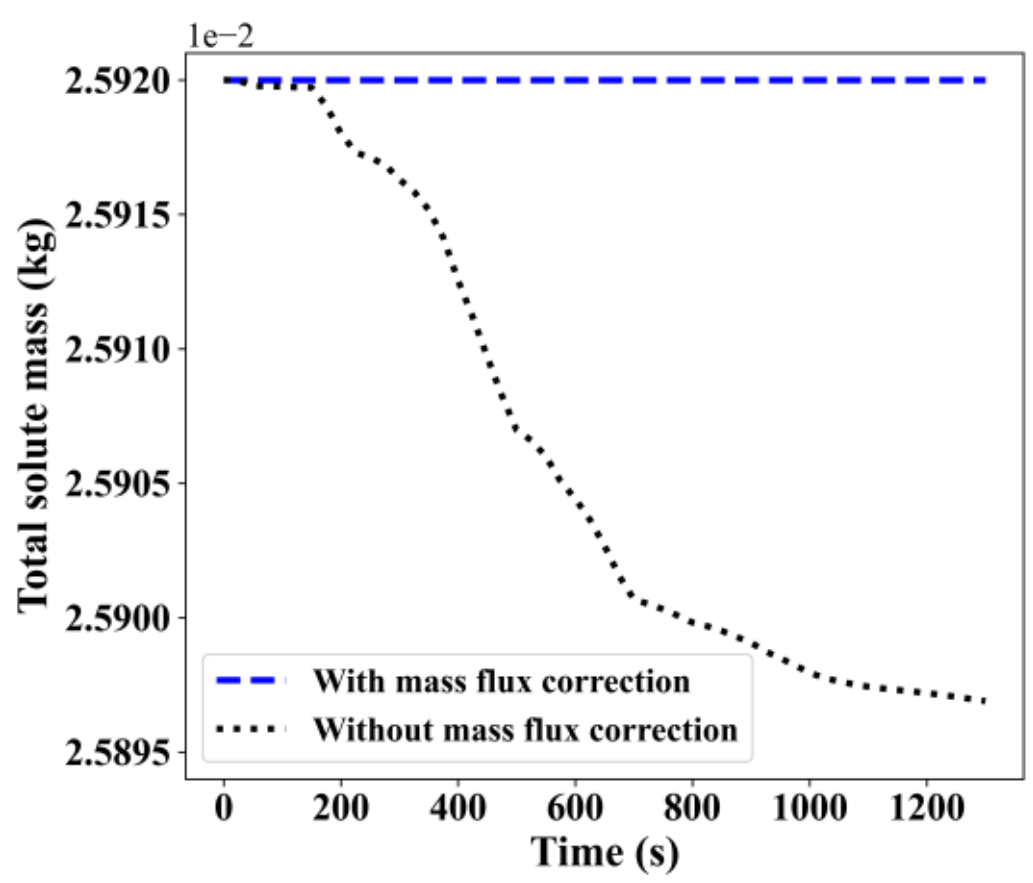

**Figure 11** Benchmark 6: temporal variation of the total solute mass in the alloy with and without mass flux correction.

*4.5. Alloy droplet-wall impact, solidification, and MS*

In this section, the complete coupled multi-physics model is employed to study the impact, solidification, and MS of the alloy droplet deposition problem (Benchmark 7) introduced in section 2.1. The Lever rule is used for the microsegregation modeling, unless stated otherwise. The predicted temporal evolution of the droplet shape upon impact and solidification is compared with the reference numerical and experimental data (Xue *et al.* 2007) in Figure 12. As illustrated by the experimental and simulation results, the droplet impact is in the spreading regime, and the final splat morphology is a disk with a pronounced rim. This demonstrates that the present simulation accurately predicts both the shape evolution during the impact and the final splat configuration, showing fine agreement with the experimental results. Based on $\mathrm{BR} = -10.4$ the thermal buoyancy force is roughly 10 times stronger than and in opposite direction to the solutal buoyancy force, however, the small Rayleigh number ($\mathrm{Ra} = \mathrm{Gr}\,\mathrm{Pr} = 370$) for this case shows a negligible natural convection effect.

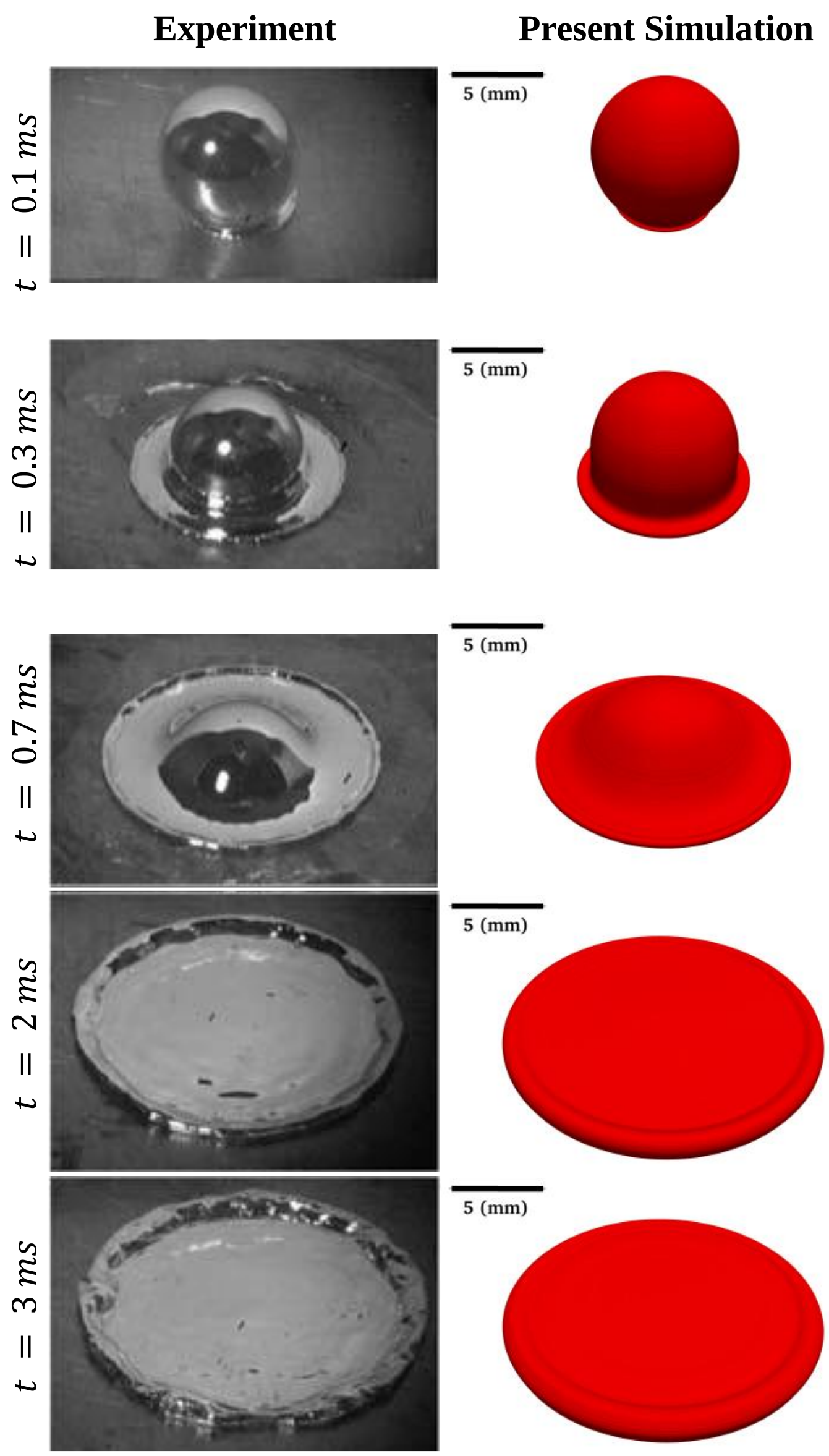

**Figure 12** Benchmark 7: the droplet shape evolution. The comparison of the present simulation with the experimental images (Xue *et al.* 2007).

Figure 13 provides a comparison of the predicted spread factor variation over time by the present simulations, the reference numerical simulation, and the empirical measurements (Xue *et al.* 2007). Two separate simulations are obtained in the current work: the default model accounting for mass (species) transport, and the conventional model without considering mass transfer. In the conventional model, the concentration transport equation is not solved, and the solid mass fraction is obtained by a simple relation:

$$f_s = \begin{cases} 0 & ;\ T \geq T_{\text{liq}}^0 \\ 1 & ;\ T \leq T_{\text{sol}}^0 \\ \frac{T_{\text{liq}}^0 - T}{T_{\text{liq}}^0 - T_{\text{sol}}^0} & ;\text{otherwise} \end{cases}, \tag{44}$$

where $T_{\text{liq}}^0 = 877.15\ K$ and $T_{\text{sol}}^0 = 850\ K$ are the liquidus and solidus temperatures at the nominal AlA380 alloy composition. Note that the reference numerical simulation (Xue *et al.* 2007) used an extra simplification, which is considering the alloy as a pure material with a melting temperature of $T_{\text{melt}} = 843\ K$. This assumption also makes the use of Eq. (44) with $T_{\text{liq}}^0 = T_{\text{melt}} + \Delta T$ and $T_{\text{sol}}^0 = T_{\text{melt}} - \Delta T$, where $\Delta T$ is a small temperature difference of the order of $1\ K$.

As shown in Figure 13, even the present model without mass transfer, accounting for an approximate mushy region temperature span by Eq. (44), improves the prediction, compared to the reference numerical simulation (Xue *et al.* 2007) assuming a pure material phase change. The inclusion of mass transfer, i.e., the present default model, further enhances the prediction, especially for the maximum and final spread factors. The relative error of the maximum spread factor prediction with respect to the experimental data is 7% for the present default model, 14% for the present model without mass transfer, and 17% for the reference simulation (Xue *et al.* 2007). The remaining error in spread factor prediction by the best (default) model may be potentially attributable to the absence of contact angle hysteresis information and the effects of multi-component alloy species transfer (vs. the two-component alloy considered here).

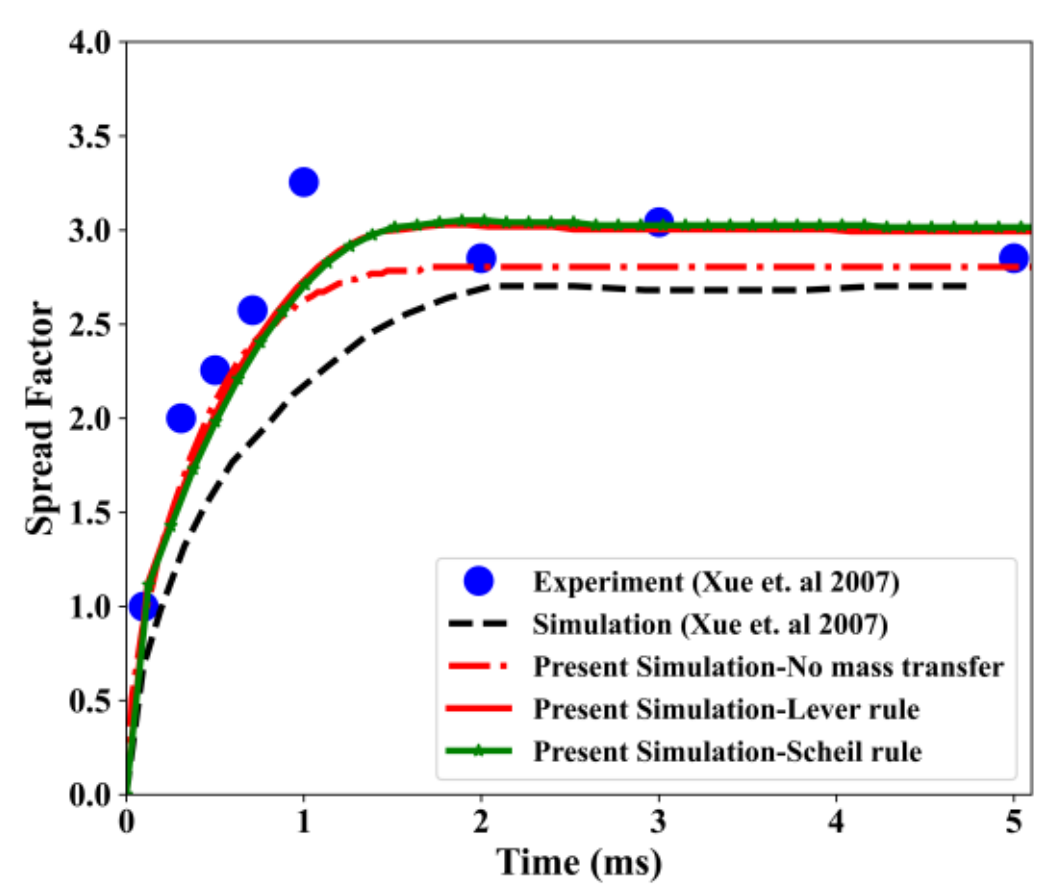

**Figure 13** Benchmark 7: The spread factor versus time. Present simulations with mass transfer (and Lever or Scheil microsegregation rule) and without mass transfer effects against the reference numerical solution (without mass transfer) and experimental measurements (Xue *et al.* 2007).

The present results suggest that the mass transfer phenomenon can affect the dynamics of an alloy droplet impact and solidification, the effect which has always been neglected in the literature. To analyze the reason behind this observation, we first compare the present model without mass transfer and the reference numerical simulation by Xue *et al.* (2007). The main distinction is the temperature difference across the mushy region, which is $T_{\text{liq}} - T_{\text{sol}} = 27.15\ K$ for the former and about $1\ K$ for the latter model. This difference brings about a much thicker mushy region for the former model. The extent of the mushy region influences the resistance of the melt flow to spread over surface. In the default model, accounting for mass transfer, the thickness of the mushy region can be different from $27.15\ K$, since this temperature difference is not constant and varies by time and space, as a result of the variation of the liquidus and solidus temperatures by the local solute concentration, see Eqs. (26) and (27). To support this discussion, a comparison of the mushy region configuration, indicated by the solid fraction iso-lines, and the contours of the radial component of the momentum loss term, $\boldsymbol{\nabla} \cdot [\mu(\boldsymbol{\nabla u} + (\boldsymbol{\nabla u})^T)] + \boldsymbol{S}_{m,u}$, in Eq. (11) for the simulations with and without mass transfer at two time instances before the peak spread factor is presented in Figure 14. Note that the contribution of the porous medium pressure loss, i.e., $\boldsymbol{S}_{m,u}$, is dominant in this term. As can be observed in this figure, the inclusion of mass transfer in the model led to a wider mushy region, especially near the advancing rim of the splat, compared to the model without mass transfer with constant $T_{\text{liq}} - T_{\text{sol}} = 27.15\ K$. Based on Eqs. (26) and (27), the temperature difference across the mushy region equals

$$T_{\text{liq}} - T_{\text{sol}} = \begin{cases} m_{\text{liq}} C_1 (K_p - 1)/K_p & ; C_1 \leq K_p C_{\text{eut}} \\ (T_{\text{melt}} - T_{\text{eut}}) + m_{\text{liq}} C_1 & ; C_1 > K_p C_{\text{eut}} \end{cases}. \tag{45}$$

Based on Eq. (45) and the phase diagram data given in Appendix A, $T_{\text{liq}} - T_{\text{sol}} > T_{\text{liq}}^0 - T_{\text{sol}}^0 = 27.15\ K$ where the local concentration lies within the range $0.00618 < C_1 < C_0 = 0.085$. The mushy zone at the early stages of solidification, shown in Figure 14 ($t = 0.8\ ms$, and $1.0\ ms$), mainly lies within the solute-lean region with $C_1 < C_0$ (or a negative segregation degree) which increases the temperature span compared to the nominal difference, $T_{\text{liq}}^0 - T_{\text{sol}}^0 = 27.15\ K$. This will be supported further by examining the segregation map at early stages of solidification in Figure 15. The increase in this temperature span justifies why the mushy region expands by accounting for mass transfer. The wider mushy region results in the melt flow velocity changes to zero in a wider region, inducing less intense momentum loss, and the splat experiences reduced resistance to outward radial motion, promoting later arrest and larger spreading.

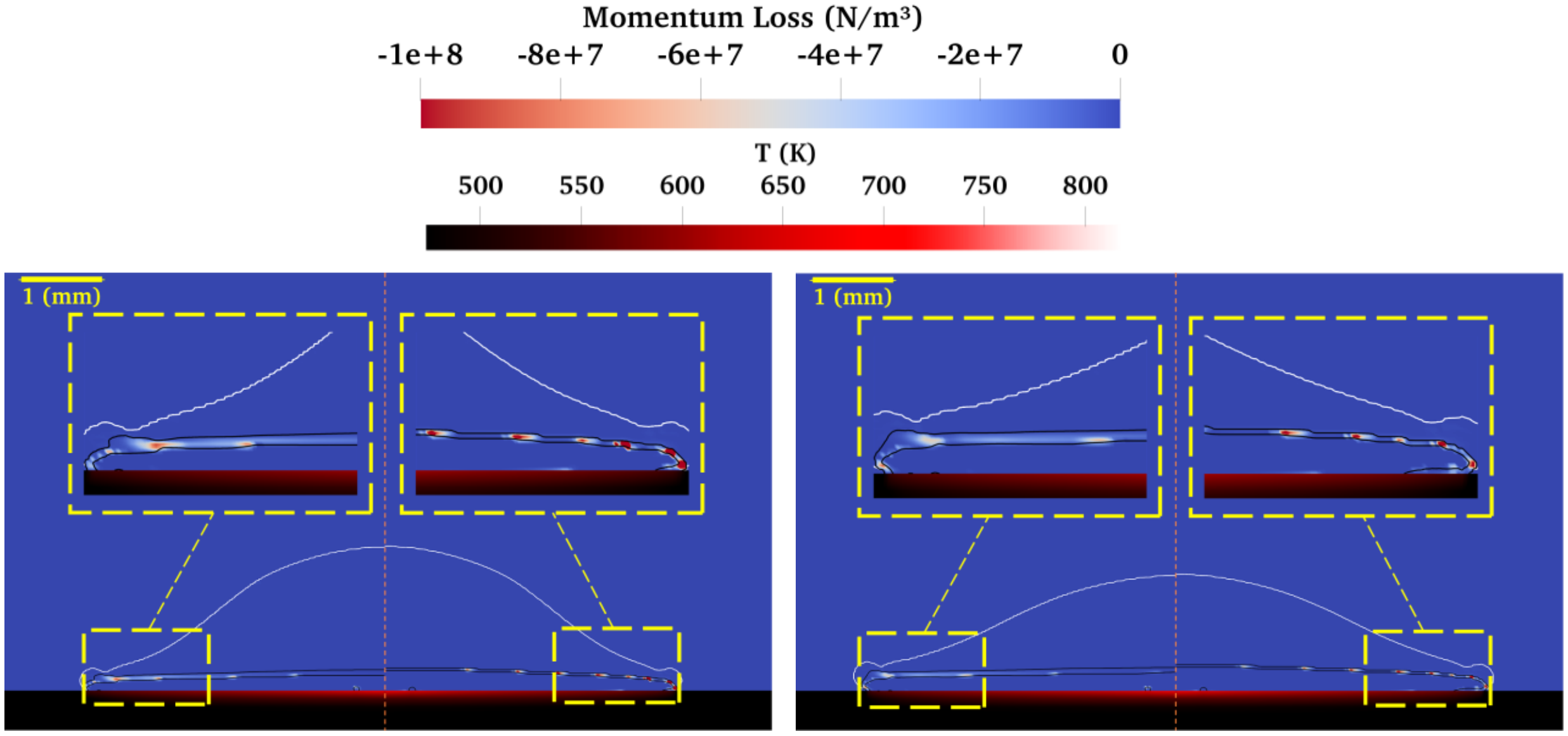

**(a)** **(b)**

**Figure 14** Benchmark 7: contours of the radial component of the momentum loss term, $\boldsymbol{\nabla} \cdot [\mu(\boldsymbol{\nabla u} + (\boldsymbol{\nabla u})^T)] + \boldsymbol{S}_{m,u}$, along with temperature contours in the substrate, for the present simulations. The left and right side of the dashed line in each panel shows the result with and without mass transfer effects, respectively. The white solid line represents the droplet interface ($\alpha_1 = 0.5$), and the black solid lines indicate the mushy region borders ($f_s = 0.95$ and $f_s = 0.05$). a) $t = 0.8\ ms$, and b) $t = 1.0\ ms$.

As a quantitative measure of flow resistance, the overall rate of the radial momentum loss within the droplet is computed by

$$\int_D \alpha_1 \{\boldsymbol{\nabla} \cdot [\mu(\boldsymbol{\nabla u} + (\boldsymbol{\nabla u})^T)] + \boldsymbol{S}_{m,u}\}_r dV, \tag{46}$$

where the integration is performed over the entire domain, and subscript $r$ denotes the radial component of a vector. The values of this quantity for the default case (with mass transfer) and the simplified case (without mass transfer) are $-0.1815\ N$ and $-0.2509\ N$, respectively, which further support the aforementioned discussion.

Therefore, the inclusion of more physics by including mass transfer and concentration-dependent solidus and liquidus temperatures can bring about more accurate droplet spread factor and impact dynamics prediction. This comparison reveals that the mass transfer phenomenon can have important effects on the dynamics of a spreading droplet. The more important advantage of considering mass transfer is the possibility to predict concentration non-uniformity and MS map. Figure 15 illustrates the progression of solidification through a set of complementary visualizations, including contours of the segregation degree, the solid mass fraction with superimposed velocity vectors, and the temperature field within the solid substrate. Solidification initiates at the droplet-substrate contact area, see Figure 15a. The solidification front entraps a small air pocket stuck beneath the splat near this contact area, making a kind of macro-shrinkage or porosity defect depending on the volume of the entrapped bubble(s). This bubble is a portion of the draining air film trapped beneath the droplet due to viscous resistance of the draining air, during the early stages of the impact, and is known as the "impact bubble" in the literature (Hendrix et al. 2016). The solidification front subsequently propagates outward, as the lamella advances on the substrate, and upward into the droplet, which entraps the impact bubble within the solidified layer as shown in the inset of Figure 15a. Previous studies (Yi *et al.* 2022, Wu *et al.* 2026) have shown that, during solidifying droplet impact, entrapped gas pockets act as important precursors of defect formation, such as internal porosity, interfacial voids, and lack-of-fusion, in the deposited material. This occurs because rapid solidification prevents the entrapped air from escaping. As a quantitative measure of porosity, the final macro-shrinkage parameter for our case is computed to be $\mathrm{MSh}_{\infty} = 0.422\ \%$.

The solute rejected at the solidification front within the lamella is advected away toward the rim by the liquid flow. This flow is intensified within the mushy region between $t = 0.5\ ms$ and $t = 1.2\ ms$. This strong advection washes away the solute-rich melt in the mushy region and leaves a vein of negatively segregated solid within the lamella after the solidification (Figure 15b and c). At $t = 1.2\ ms$, the solidification of a portion of the rim, adjacent to the substrate, slows down its spread. This makes the liquid part of the rim lift off from the substrate (Figure 15c) and roll down subsequently (Figure 15d), encapsulating a pocket of air beneath the spreading splat. This air entrapment mechanism is different from the two known mechanisms reported in hydrodynamics impacts, i.e., the "impact bubble" (Hendrix et al. 2016) and "entrapped bubble" (Pittoni *et al.* 2015) mechanisms, and is observed in the current work for the first time. We call this mechanism "rolling bubble", which can be seen in impacts with solidification. Comparing the solidification front between $t = 1.2\ ms$ and $t = 1.4\ ms$, it is revealed that a severe remelting occurs at a portion of the front where the lamella attaches to rim (Figure 15d) (the remelting can be more clearly observed in the supplementary video), leading to a hook formation in this region. This hook is clearly visible in the mushy region at $t = 1.6\ ms$ (Figure 15e). The formation of this hook has an important influence on MS. The hook traps a pocket of melt which produces a solute-rich region, generating a positively segregated spot within the hook structure, as can be seen in Figure 15d and e. Hook formation and solute advection are two tightly coupled mechanisms that control whether a region becomes positively or negatively segregated during solidification. Hook formation occurs when insufficient superheat or strong thermal gradients cause early solidification, creating a curved solidified structure (the "hook") that partially traps liquid steel and modifies local flow paths. This trapped and redirected liquid becomes a preferential site for solute redistribution because it alters solid–liquid interface stability (Thomas 2017). At the same time, solute advection continuously transports enriched or depleted melt through these altered flow paths. The combined effect determines the spatial pattern of segregation by controlling where solute-rich or solute-lean liquid accumulates. In regions where solute is swept away by strong washing flow or redirected outward by hook-induced flow barriers, negative segregation develops due to local solute depletion (Zhang *et al.* 2021, Chen *et al.* 2022). Conversely, in zones where advection brings solute-enriched liquid into confined regions, such as beneath hooks, positive segregation forms due to solute accumulation and delayed solidification (Takeuchi & Brimacombe 1985). This latter mechanism leads to the formation of positively segregated spot within the hook in Figure 15d and e.

As the flow advection washes down the mushy region, it redistributes solute away from the splat central axis toward the rim and accumulates the solute-rich melt at the tip of the rim. At $t = 2.3\ ms$ (Figure 15f), the solidification front touches the upper surface of the splat, separating the melt into two distinct regions: the solute-rich rim region and the central splat top surface region. The completion of solidification in the former region produces another positively segregated zone at the tip of the rim (Figure 15g).

The final MS map at the end of the solidification is shown in Figure 15g. The reasons behind the formation of different regions on this map were discussed before. Note that there is another (moderately) positively segregated region in the center of the solidified splat which starts forming around $t = 1.6\ ms$ (Figure 15e). This zone appears due to the formation of a stagnation point near the center of the spreading droplet, promoting the accumulation of rejected solute in this region. In general, the rim of the lamella shows the most pronounced positive segregation within the solidified droplet, followed by the area within the hook structure formed during the solidification. In contrast, the region within the solidified lamella develops a band of negative segregation, which is extended diagonally toward the center of the splat and near the substrate surface. These compositional inhomogeneities have significant impacts on the susceptibility to defect formation, such as cracking and other solidification-related imperfections.

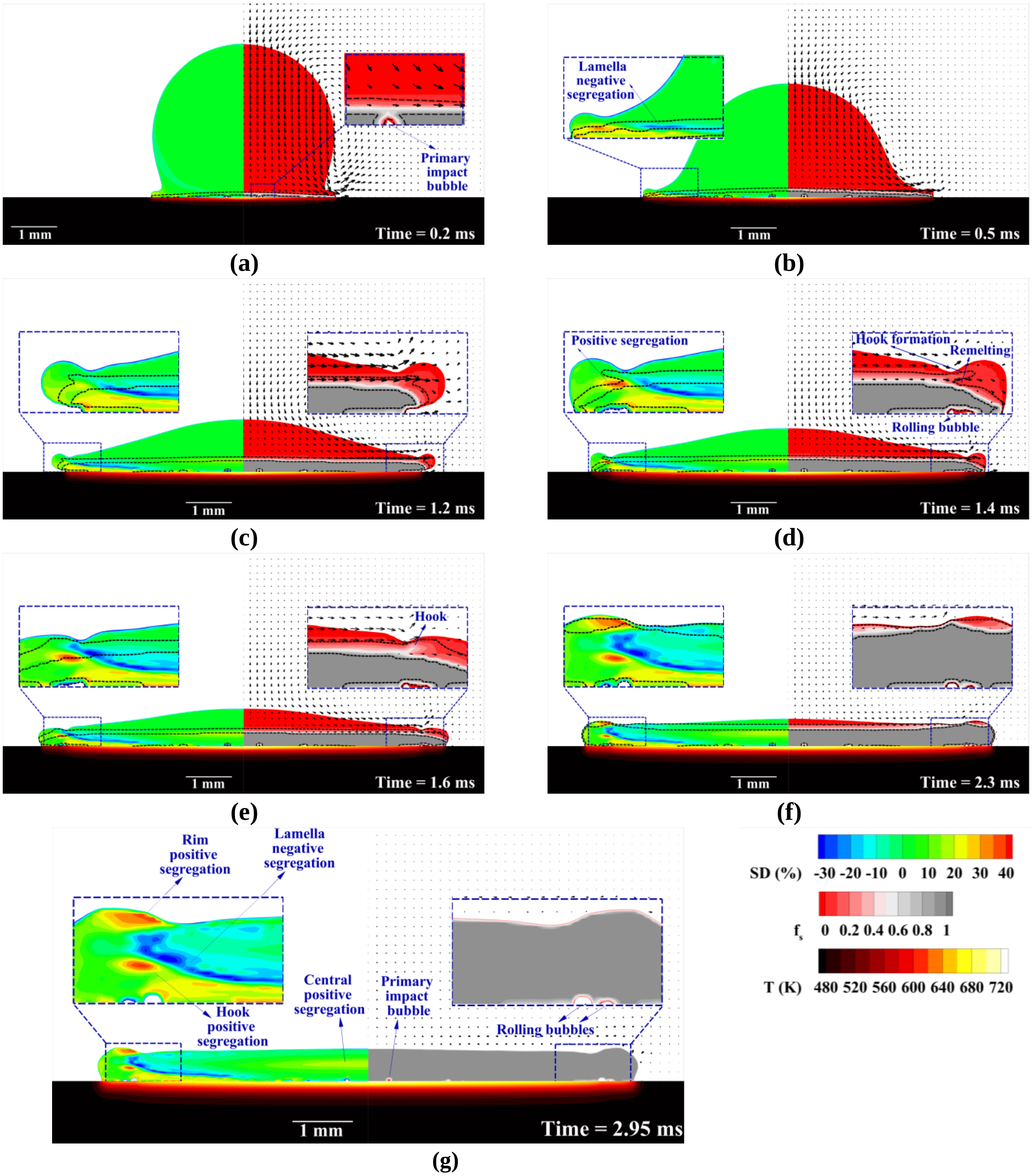

**Figure 15** Benchmark 7: The left and right side of each panel shows the contours of solid mass fraction ($f_s$), with superimposed velocity vectors, and segregation degree (SD), respectively, with contours of temperature within the substrate. The black dashed lines in a-f indicate the mushy region borders ($f_s = 0.95$ and $f_s = 0.05$). a) $t = 0.2\ ms$, b) $t = 0.5\ ms$ c) $t = 1.2\ ms$, d) $t = 1.4\ ms$, e) $t = 1.6\ ms$, f) $t = 2.3\ ms$, and g) $t = 2.95\ ms$ (complete solidification).

In Figure 15, we used the Lever rule for microsegregation which assumes local (subgrid scale) thermodynamic equilibrium at the solid-liquid interface. To examine the effect of this assumption on MS in the present problem, an additional simulation is performed using the Scheil rule. In practice, the actual redistribution of solute during solidification lies between the two extremes represented by the Lever and Scheil rules. Figure 16a compares the MS maps predicted by the two microsegregation models. Overall, both models predict nearly identical final splat shapes, primary impact and rolling bubbles, and similar SD patterns and critical regions, indicating that macroscopic effects play a dominant role in determining the MS map. According to Figure 13, the splat spreading dynamics is also insensitive to the choice of the microsegregation model. However, some secondary differences can also be observed in MS maps (Figure 16a). There exist oscillatory patterns and moderate discontinuities in the negative segregation regions predicted by the Scheil rule, which may be due to physical or even numerical instabilities. The overall maximum SDs are $\mathrm{SD}_{\infty,\max} = +45.6\%$ for the Lever and $+47.2\%$ for the Scheil rule. For a more quantitative comparison, the SD profiles along a section of the splat are compared in Figure 16b. While the Lever and Scheil models show similar MS trends, Scheil predicts a slightly stronger positive segregation intensity, about 18.8% larger negative segregation than the Lever model within a small region (the local minimum of SD are $\mathrm{SD}_{\infty,\min} = -32.2\%$ for the Lever and $-51.0\%$ for the Scheil rule), and additional SD fluctuations appearing in the region $3 < r < 3.7$ mm near the rim of the lamella. This observation is aligned with the one reported by Dong *et al.* (2017) for MS in billet continuous casting, where the Scheil model generally predicted higher SD values in positively segregated regions and lower SD values in negatively segregated zones. Despite these differences, the global segregation structure is largely insensitive to the choice of the microsegregation model for the present case, suggesting reasonable validity of the local equilibrium assumption for identifying the location of critical segregation zones.

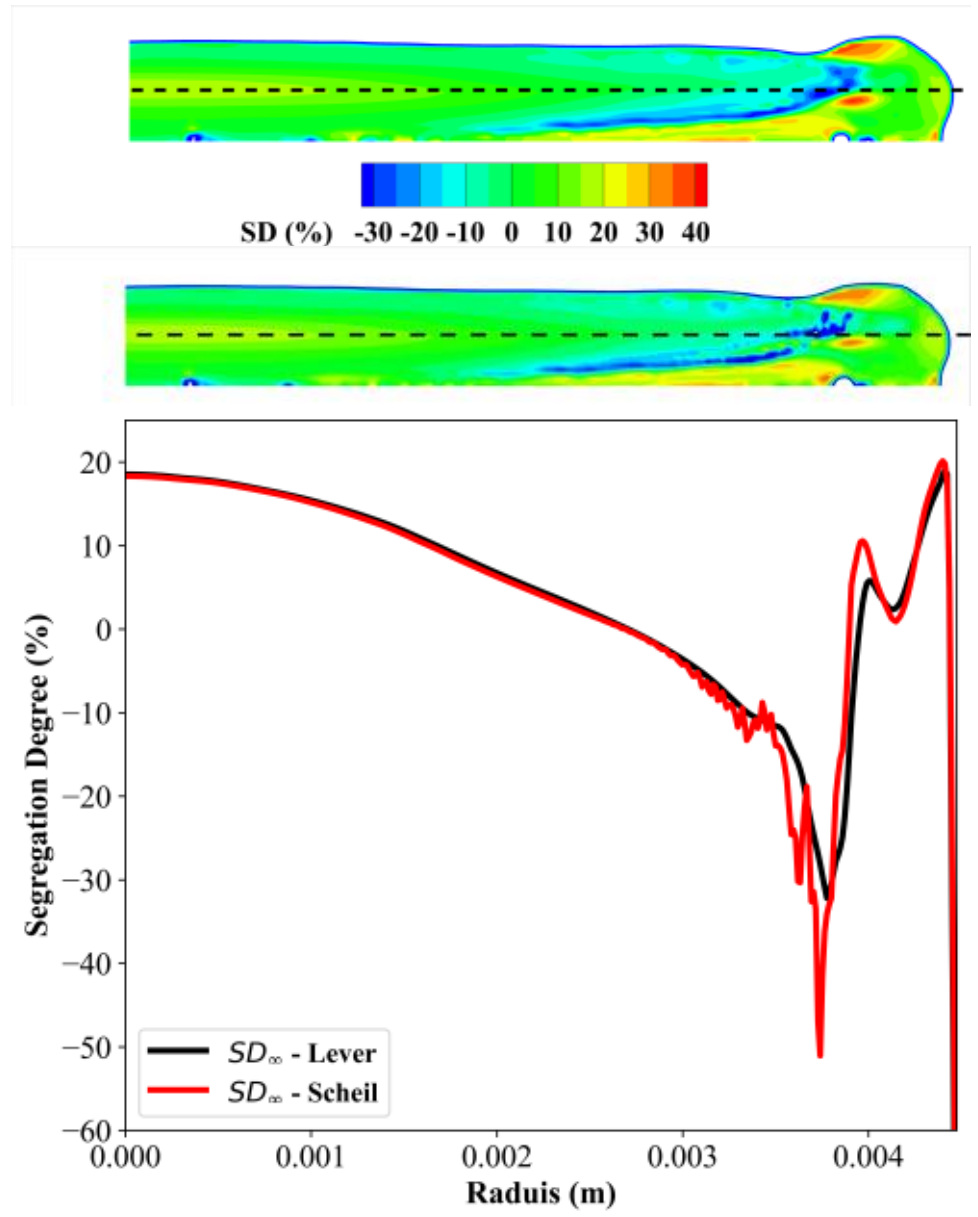

**Figure 16** Benchmark 7: Comparisons of the segregation degrees at $t = 2.95\ ms$ (complete solidification) predicted by the Lever and Scheil rules. The contours (top), and the radial profiles along the dashed line (bottom), 0.275 mm above the substrate surface, shown in top panel.

## 5. Conclusion

In this study, a numerical framework, involving fully-coupled interactions of hydrodynamics, wetting, conjugate heat transfer, phase change, and mass transfer, was developed to model impact, solidification, and MS of a metallic alloy droplet. The model predictive capability was established through a rigorous, stepwise validation strategy comprising a hierarchical set of benchmark problems, spanning droplet impact hydrodynamics, pure-metal heat transfer and solidification, immiscible scalar transport in two-phase systems, and MS during alloy solidification in a mold. Application of the framework to the impact and solidification of an AlA380 alloy droplet demonstrated that solute transport plays a decisive role in governing both transient dynamics and final solidification outcomes. Accounting for mass transfer led to reduced localization of momentum loss within the mushy region and consequently increased radial spreading, compared with simulations neglecting mass transfer. Further analysis of the results demonstrated that this is because the dynamically evolving solute field modifies local liquidus and solidus temperatures, influencing solidification kinetics and the thickness of the mush. In addition, a new mechanism of air entrapment specific to solidifying impacts was observed, which arose due to the lift off of the advancing rim from the substrate and its subsequent rolling down, triggered by the attachment of the solidifying bottom layer of the splat to the substrate. This mechanism, named “rolling bubble”, contributes to the macro-shrinkage and porosity of the final solidified object. Finally, the analysis of the MS map revealed that the ultimate solute distribution is strongly spatially non-uniform and is governed by the interaction between impact-induced flow structures and possible remelting and hook formation during solidification. In particular, the splat rim consistently exhibits the highest degree of positive segregation, resulting from solute-rich melt accumulation sustained by solute rejection at the advancing solidification front combined with outward radial advective flow within the mush. These findings highlight the importance of mass transfer modeling for predictive assessment of spreading behavior and segregation-driven defect formation in droplet-based metal additive manufacturing.

The Weber number, primarily characterizing the impact velocity, is a key parameter governing droplet impact dynamics. Excessively high Weber numbers lead to violent spreading, droplet breakup, gas entrapment, and splashing, which can intensify defect formation (Yuan *et al.* 2019, Li *et al.* 2026). In some systems, critical Weber thresholds define transitions between deposition, rebound, and disintegration, while solidification times and final morphologies also depend strongly on Weber number (Gielen *et al.* 2019, Chowdhury *et al.* 2024). Conversely, a too low Weber number may result in insufficient spreading and incomplete fusion. A thorough parametric study on the effect of Weber number as well as other design parameters defined in Table 1 on MS in 3D printing, using the present numerical model, would be an interesting topic for future studies. The other future directions could be the extension of the present research to consecutive droplet deposition, evaluating the residual stress (by adding the thermomechanical equations to the model), and/or accounting for multicomponent transport.

**Declaration of Interests:** The authors report no conflict of interest.

## Supplementary Material

The supplementary material provides a text file including thermophysical properties and phase diagram data of the materials used in all benchmarks, an OpenFOAM case including complete case setup of the alloy droplet case, and videos for the animated impact, solidification, and MS of the alloy droplet case.

## Data availability

The data that support the findings of this study are available from the corresponding author upon reasonable request.

## Appendix A: Material properties

Table 4 and Table 5 provide the thermophysical properties and phase diagram information for the main case: AlA380 alloy droplet, air, and steel substrate. The details of the other benchmarks' data are reported in the supplementary material.

**Table 4** The material properties and phase-diagram data for AlA380.

| Parameters | Value | Reference |
|---|---|---|
| Liquid density, $\rho_l$ $(kgm^{-3})$ | 2570 | (Xue *et al.* 2007) |
| Solid density, $\rho_s$ $(kgm^{-3})$ | 2570 | |
| Liquid dynamic viscosity, $\mu_l$ $(Pa.s)$ | $1.1565 \times 10^{-3}$ | |
| Liquid conductivity, $k_l$ $(Wm^{-1}K^{-1})$ | 70 | |
| Solid conductivity, $k_s$ $(Wm^{-1}K^{-1})$ | 147.5 | |
| Liquid heat capacity, $c_{p,l}$ $(Jkg^{-1}K^{-1})$ | 1000 | |
| Solid heat capacity, $c_{p,s}$ $(Jkg^{-1}K^{-1})$ | 1050 | |
| Surface tension coefficient, $\sigma$ $(Nm^{-1})$ | 1.07 | |
| Latent heat of solidification, $L_f$ $(Jkg^{-1})$ | 389000 | |
| Nominal solute concentration, $C_0$ $(-)$ | 0.085 | |
| Eutectic concentration, $C_{\text{eut}}$ $(-)$ | 0.126 | (Karakaya 2001) |
| Partition coefficient, $K_p$ $(-)$ | 0.131 | |
| Liquidus line slope, $m_{liq}$ $(K)$ | $-662.3$ | |
| Melting temperature, $T_{\text{melt}}$ $(K)$ | 933.45 | |
| Eutectic temperature, $T_{\text{eut}}$ $(K)$ | 850 | |
| Initial solidus temperature, $T_{\text{sol}}^0$ $(K)$ | 850 | |
| Initial liquidus temperature, $T_{\text{liq}}^0$ $(K)$ | 877.15 | |
| Diffusion coefficient in solid, $D_s$ $(m^2s^{-1})$ | $4.2 \times 10^{-12}$ | (Ghods *et al.* 2016) |
| Diffusion coefficient in liquid, $D_l$ $(m^2s^{-1})$ | $4.2 \times 10^{-9}$ | |
| Solute expansion coefficient, $\beta_{C,l}$ $(-)$ | $1.31 \times 10^{-3}$ | |
| Thermal expansion coefficient, $\beta_{T,l}$ $(K^{-1})$ | $1.45 \times 10^{-4}$ | (Mills 2002) |
| Dendrite secondary arm spacing, $\lambda_2$ $(m)$ | $76 \times 10^{-6}$ | (Liu *et al.* 2015) |

**Table 5** The steel substrate and air material properties (Xue *et al.* 2007).

| Parameters | Steel $(k = w)$ | Air $(k = g)$ |
|---|---|---|
| Density, $\rho_k$ $(kgm^{-3})$ | 7900 | 1.184 |
| Dynamic viscosity, $\mu_k$ $(Pa.s)$ | - | $1.85 \times 10^{-5}$ |
| Conductivity, $k_k$ $(Wm^{-1}K^{-1})$ | 24 | 0.02551 |
| Heat capacity, $c_{p,k}$ $(Jkg^{-1}K^{-1})$ | 600 | 1005 |

## Appendix B: The grid-independence check

To ensure the fidelity of the numerical results and to minimize mesh-induced uncertainties, a systematic grid independence study was performed. The spatial discretization was varied through CPR, which controls the mesh resolution. A total of seven grids, ranging from CPR = 40 to 100, were examined. Key hydrodynamic, thermal, and solutal metrics were monitored to evaluate numerical convergence. The relative error of each objective was computed with respect to the values obtained for the finest resolution (CPR = 100) and is illustrated in Figure 17. According to this figure, the final maximum segregation degree, $SD_{\infty,max}$, and the final minimum segregation degree, $SD_{\infty,min}$, governed by mass transfer length scales require much finer grid resolutions to achieve grid-independent results, compared to the other objectives. The errors of all objective variables are below 1% for CPR $\geq$ 90.

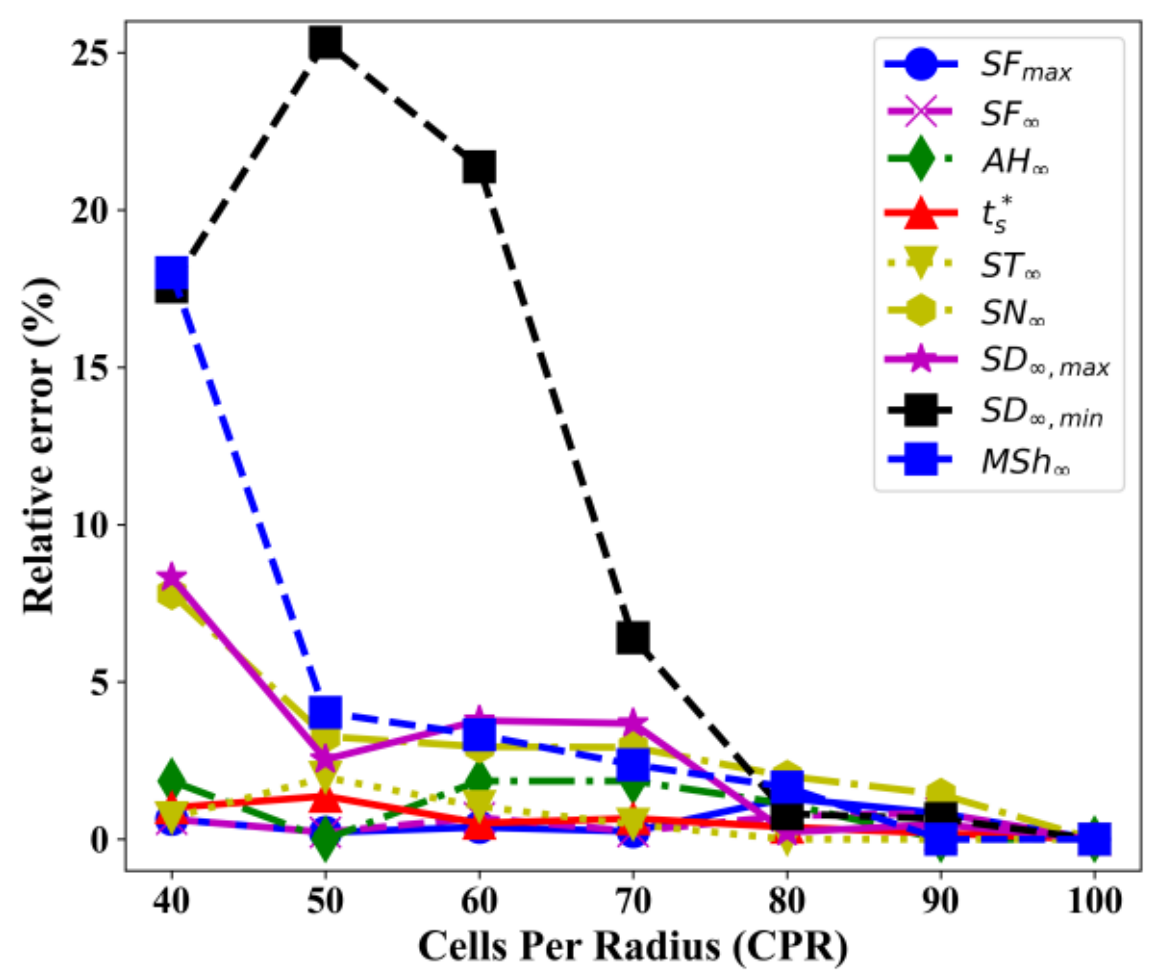

**Figure 17** The grid-independence results: the relative errors for different objectives versus the grid resolution.

For an additional assessment, a Grid Convergence Index (GCI) analysis (Celik *et al.* 2008) is also conducted for the most sensitive objectives, i.e., $SD_{\infty,max}$ and $SD_{\infty,min}$, in Table 6. These two extrema characterize the intensity of MS within the solidified droplet and are governed by the strongly coupled interaction of momentum transport, thermal gradients, solute advection–diffusion, and phase change in the mushy zone. Because steep compositional gradients develop near the advancing solidification front, these metrics provide a stringent test of spatial resolution adequacy. According to this data, the mean uncertainties of SD computation ($GCI^{21}$) are below 1%.

**Table 6** The GCI index for SD computation.

| Parameter | $\emptyset = \mathbf{SD_{max}}$ Values | $\emptyset = \mathbf{SD_{min}}$ Values |
|---|---|---|
| $(\mathrm{CPR})_1, (\mathrm{CPR})_2, (\mathrm{CPR})_3$ | 100, 90, 80 | 100, 90, 80 |
| $r_{32}, r_{21}$ | 1.11, 1.13 | 1.11, 1.13 |
| $\emptyset_1, \emptyset_2, \emptyset_3$ | 45.78, 45.56, 45.67 | -53.03, -52.68, -53.46 |
| $e_{\mathrm{ext}}^{21}$ | 0.42% | 0.60% |
| $\mathrm{GCI}^{21}$ | 0.53% | 0.75% |